\documentclass[aps,prl,twocolumn]{revtex4-2}

\usepackage{mathrsfs}
\usepackage{bbm}
\usepackage{amsmath}
\usepackage{amssymb}
\usepackage{amsthm}
\usepackage{graphicx}
\usepackage{MnSymbol}
\usepackage{mathtools}
\usepackage{stmaryrd}
\usepackage{bbold}
\usepackage[italicdiff]{physics}

\makeatletter

\usepackage{hyperref}
\usepackage{color}
\definecolor{myblue}{RGB}{0,50,200}
\hypersetup{
colorlinks,
citecolor=myblue,
linkcolor=myblue,
urlcolor=myblue
}

\allowdisplaybreaks

\newcommand{\mca}{\mathcal}
\newcommand{\mbb}{\mathbb}

\newcommand{\Var}[1]{{\rm var}\qty[#1]}
\newcommand{\sectionprl}[1]{{\em #1}\/.---}

\begin{document}
\title{Thermodynamics of Precision in Markovian Open Quantum Dynamics}

\author{Tan Van Vu}
\email{tanvu@rk.phys.keio.ac.jp}

\affiliation{Department of Physics, Keio University, 3-14-1 Hiyoshi, Kohoku-ku, Yokohama 223-8522, Japan}

\author{Keiji Saito}
\email{saitoh@rk.phys.keio.ac.jp}

\affiliation{Department of Physics, Keio University, 3-14-1 Hiyoshi, Kohoku-ku, Yokohama 223-8522, Japan}

\date{\today}

\begin{abstract}
The thermodynamic and kinetic uncertainty relations indicate trade-offs between the relative fluctuation of observables and thermodynamic quantities such as dissipation and dynamical activity. Although these relations have been well studied for classical systems, they remain largely unexplored in the quantum regime. In this paper, we investigate such trade-off relations for Markovian open quantum systems whose underlying dynamics are quantum jumps, such as thermal processes and quantum measurement processes. Specifically, we derive finite-time lower bounds on the relative fluctuation of both dynamical observables and their first passage times for arbitrary initial states. The bounds imply that the precision of observables is constrained not only by thermodynamic quantities but also by quantum coherence. We find that the product of the relative fluctuation and entropy production or dynamical activity is enhanced by quantum coherence in a generic class of dissipative processes of systems with nondegenerate energy levels. Our findings provide insights into the survival of the classical uncertainty relations in quantum cases.
\end{abstract}

\maketitle

\sectionprl{Introduction}Small systems are inevitably subjected to significant fluctuations owing to their interaction with the environment; these fluctuations can strongly affect the performance of physical systems such as heat engines, mechanical clocks, and molecular motors. Thus, understanding fluctuations is an important step, both theoretically and practically, to controlling or overcoming such effects.

The fluctuation theorem \cite{Evans.1993.PRL,Gallavotti.1995.PRL,Crooks.1999.PRE,Jarzynski.2000.JSP,Esposito.2009.RMP,Campisi.2011.RMP}, which encodes fluctuations of thermodynamic quantities into a universal equality, has been a prominent achievement over the last two decades. Beyond this equality, in recent years, the thermodynamic uncertainty relation (TUR), which presents a trade-off between precision and dissipation, was discovered \cite{Barato.2015.PRL,Gingrich.2016.PRL,Horowitz.2017.PRE}. Qualitatively, the TUR implies that the precision of time-integrated currents, which is quantified via the relative fluctuation, cannot be enhanced without increasing dissipation. The TUR was initially developed for steady-state Markov jump processes and subsequently generalized to arbitrary initial states \cite{Dechant.2018.JSM,Liu.2020.PRL}, given by the inequality
\begin{equation}\label{eq:org.TUR}
\frac{\Var{J}}{\ev{J}^2}\ge\frac{2(1+\delta J)^2}{\Sigma_\tau},
\end{equation}
where $\ev{J}$ and $\Var{J}$ are, respectively, the mean and variance of the current $J$, $\delta J\coloneqq\tau\partial_\tau\ln|\ev{J}/\tau|$, and $\Sigma_\tau$ denotes the irreversible entropy production during the operational time $\tau$. A similar relation that applies to arbitrary counting observables is the kinetic uncertainty relation (KUR) \cite{Garrahan.2017.PRE,Terlizzi.2019.JPA}, which is obtained by replacing the irreversible entropy production in Eq.~\eqref{eq:org.TUR} with dynamical activity. Notably, the TUR leads to a universal trade-off between the power and efficiency of heat engines \cite{Pietzonka.2018.PRL,Shiraishi.2016.PRL}. Furthermore, it can be applied to infer dissipation from trajectory data without prior knowledge of the system \cite{Li.2019.NC,Manikandan.2020.PRL,Vu.2020.PRE,Otsubo.2020.PRE}. Numerous studies have examined and extended these uncertainty relations for both classical and quantum dynamics \cite{Gingrich.2017.PRL,Brandner.2018.PRL,Agarwalla.2018.PRB,Hasegawa.2019.PRE,Saryal.2019.PRE,Vu.2019.PRE.UnderdampedTUR,Liu.2019.PRE,Hasegawa.2019.PRL,Timpanaro.2019.PRL,Horowitz.2020.NP,Vu.2020.PRR,Koyuk.2020.PRL,Potts.2019.PRE,Vo.2020.PRE,Falasco.2020.NJP,Friedman.2020.PRB,Sacchi.2021.PRE,Park.2021.PRR,Lee.2021.PRE,Dechant.2021.PRR,Timpanaro.2021.arxiv}.

\begin{table}[t]
\centering
\begin{tabular}{ll l}
\hline\hline
$\Var{\bullet}/\expval{\bullet}^2\ge\mca{B}_\bullet$ & & ~~~{Dissipative processes} \\ [0.4ex]
\hline
${\mca{B}_J=2(1+\delta J)^2/\Sigma_\tau}$ & [Eq.~\eqref{eq:main.result.1}] & ~~~Nonresonant\\
$\mca{B}_O=(1+\delta O)^2/\mca{A}_\tau $ & [Eq.~\eqref{eq:main.result.2}] & ~~~Generic\\
$\mca{B}_\tau=1/\expval{N}_\tau$ & [Eq.~\eqref{eq:main.result.3}] & ~~~Generic\\ [0.4ex]
\hline\hline
\end{tabular}
\caption{Sufficient conditions in which the classical uncertainty relations survive in the quantum regime. Equations \eqref{eq:main.result.1}, \eqref{eq:main.result.2}, and \eqref{eq:main.result.3} are quantum uncertainty relations for the currents that are odd under the time reversal operation, arbitrary observables, and first passage times, respectively. The right column displays the sufficient conditions to re-obtain the classical expressions $\mca{B}_\bullet$. ``Nonresonant'' implies the process where all jump operators have the form $L_k\propto\dyad{\epsilon_m}{\epsilon_n}$. This case occurs, for example, when the energy difference of any pair of energy levels is irrational, which is general in nonintegrable systems. ``Generic'' implies the case satisfying $[L_k,H]=\omega_kL_k$ with energy difference $\omega_k$, which is fulfilled in most thermal processes of interest.}
\label{table:summa}
\end{table}

It is now well known that quantum coherence plays an essential role in a broad class of thermodynamics, especially in the field of finite-time thermodynamics \cite{Horodecki.2013.NC,Uzdin.2015.PRX,Lostaglio.2015.PRX,Korzekwa.2016.NJP,Francica.2019.PRE,Santos.2019.npjQI,Francica.2020.PRL,Tajima.2021.PRL,Miller.2020.PRL.QLP,Vu.2022.PRL,Brandner.2017.PRL,Brandner.2020.PRL,Scully.2011.PNAS,Watanabe.2017.PRL,Menczel.2020.PRA}. Concerning the TUR and KUR, it has been shown through specific examples that these relations can be violated in the quantum realm \cite{Ptaszynski.2018.PRB,Cangemi.2020.PRB,Kalaee.2021.PRE,Menczel.2021.JPA,Bret.2021.PRE}. Despite the derivation of several quantum bounds \cite{Erker.2017.PRX,Guarnieri.2019.PRR,Carollo.2019.PRL,Hasegawa.2020.PRL,Hasegawa.2021.PRL,Miller.2021.PRL.TUR,Hasegawa.2021.PRL2}, the interplay between dissipation and quantum coherence in constraining finite-time fluctuations remains unclear. Moreover, a recent study \cite{Kalaee.2021.PRE} has shown that quantum coherence responsible for the TUR violations cannot be characterized solely by off-diagonal elements of the density matrix. Note here that quantum coherence has no unique definition \cite{Streltsov.2017.RMP}. Therefore, these backgrounds strongly motivate us to clarify what types of quantum coherence are relevant to the TUR and KUR. Elucidating this should also give us insights into when and how the original uncertainty relations survive in the quantum realm.

In this paper, we investigate the precision of observables in open quantum systems and the role of quantum coherence. Focusing on the dynamical class of quantum jump processes, we find the relevant quantum coherence terms and derive fundamental bounds on the relative fluctuation of observables in terms of both thermodynamic quantities and quantum coherence terms [cf.~Eqs.~\eqref{eq:main.result.1} and \eqref{eq:main.result.2}].
The coherence terms can help clarify whether quantum coherence reduces or enhances the relative fluctuation of observables. In addition to dynamical observables, we also derive a lower bound on the first passage time (FPT) fluctuation of generic counting observables [cf.~Eq.~\eqref{eq:main.result.3}], which is relevant when a physical observable appears in the quantum jump processes. The obtained bounds are general and valid for arbitrary operational times and initial states.
We also provide sufficient conditions to reduce these quantum generalizations of the TUR and KUR to the classical expressions, as summarized in Table \ref{table:summa}.
We numerically illustrate our findings using a three-level maser \cite{Scovil.1959.PRL,Zou.2017.PRL,Klatzow.2019.PRL}.

\sectionprl{Model}We consider an open quantum system with a finite dimension $d$, which is weakly coupled to single or multiple heat baths at different temperatures.
Let $\varrho_t$ denote the density operator of the system at time $t$. Then, its time evolution during the operational time $\tau$ can be described by the Lindblad master equation \cite{Lindblad.1976.CMP,Gorini.1976.JMP}:
\begin{equation}\label{eq:Lindblad.dyn}
\dot{\varrho}_t=\mca{L}(\varrho_t)\coloneq-i\comm{H}{\varrho_t}+\sum_{k=1}^K\mca{D}[L_k]\varrho_t. 
\end{equation}
Here, $H$ is a time-independent Hamiltonian, $\mca{D}[L]\varrho=L\varrho L^\dagger-\acomm{L^\dagger L}{\varrho}/2$ is a dissipator, and $L_k$ is the $k$th jump operator.
The dot $\cdot$ denotes the time derivative, and $\comm{\circ}{\star}$ and $\acomm{\circ}{\star}$ are the commutator and anticommutator of the two operators, respectively.
Throughout this paper, both the Planck constant and Boltzmann constant are set to unity, $\hbar=k_{\rm B}=1$. We note that the dynamics \eqref{eq:Lindblad.dyn} can describe the quantum measurement process as well as the thermal dissipation dynamics. In the quantum measurement process, $L_k$ can either represent a projective measurement on the system or characterize a jump outcome induced by continuously monitoring the environment \cite{Wiseman.2009}. For thermal dissipation dynamics, we assume the local detailed balance condition $L_k=e^{\Delta s_k/2}L_{k'}^\dagger$, which is satisfied in most cases of physical interest \cite{Manzano.2019.PRL}. Here, the operator $L_{k'}$ represents the reversed jump of the $k$th jump, and $\Delta s_k$ denotes the entropy change of the environment due to the jump.
For simplicity, we exclusively focus on time-independent driving; nevertheless, the generalization to the time-dependent case is straightforward.

The quantum dynamics \eqref{eq:Lindblad.dyn} can be unraveled into quantum jump trajectories \cite{Horowitz.2012.PRE,Horowitz.2013.NJP,Manzano.2015.PRE,Miller.2021.PRE}. For a small time interval $dt$, the Lindblad dynamics $\varrho_{t+dt}=(\mbb{1}+\mca{L}\,dt)\varrho_t$ can be expressed in the Kraus representation $\varrho_{t+dt}=\sum_{k=0}^KV_k\varrho_tV_k^\dagger$ with the operators given by
\begin{align}
V_0&\coloneq \mbb{1}-iH_{\rm eff}dt,\\
V_k&\coloneq L_k\sqrt{dt}~(1\le k\le K).
\end{align}
Here, $H_{\rm eff}\coloneqq H-(i/2)\sum_kL_k^\dagger L_k$ is the non-Hermitian Hamiltonian, and $\mbb{1}$ denotes the identity operator.
The operator $V_0$ induces a smooth nonunitary evolution, whereas operators $V_k~(k\ge 1)$ induce jumps in the system state.
Using this representation, the master equation can be unraveled into individual trajectories consisting of smooth evolution of the pure state $\ket{\psi_t}$ and discontinuous changes caused by quantum jumps in the pure state at random times.
Notably, the entire time evolution of the pure state can be described by the stochastic Schr{\"o}dinger equation \cite{Breuer.2002}.

The system is initially in a pure state $\ket{n}$ with probability $p_n$, which is confirmed by a projective measurement, that is, $\varrho_0=\sum_{n=1}^dp_n\dyad{n}$. Let $\Gamma_{\tau}=\qty{\ket{\psi_t}}_{0\le t\le\tau}$ be a stochastic trajectory of period $\tau$. Then, $\Gamma_{\tau}$ can be characterized by a discrete set $\{(t_0,n),(t_1,k_1),\dots,(t_N,k_N)\}$, where $n$ is the measurement outcome at the initial time $t_0=0$, and the $k_j$th jump  occurs at time $t_j$ for each $1\le j\le N$. By defining the time propagation operator $U(t',t)\coloneqq\exp[-iH_{\rm eff}(t'-t)]$, the probability density of observing the trajectory $\Gamma_\tau$ is calculated as
\begin{equation}
p(\Gamma_\tau)=p_n|U(\tau,t_N)\prod_{j=1}^{N}L_{k_j}U(t_j,t_{j-1})\ket{n}|^2.
\end{equation}

We consider a generic time-integrated counting observable $\Phi(\Gamma_\tau)$ defined for each trajectory $\Gamma_\tau$ as
\begin{equation}
\Phi(\Gamma_\tau)\coloneqq\sum_{j=1}^Nw_{k_j},
\end{equation}
where $w_k$ is an arbitrary real coefficient associated with the $k$th jump. In the case where the coefficients are time-antisymmetric (i.e., $w_k=-w_{k'}$ for all $k$ and its reversed counterpart $k'$), $\Phi$ is called a {\em current}. Examples of currents include the net number of jumps by setting $w_k=1=-w_{k'}$ and the entropy flux to the environment by setting $w_k=\Delta s_k=-w_{k'}$. Another observable of interest is the static observable, which is defined as \cite{fnt0}
\begin{equation}\label{eq:static.obs}
\Lambda(\Gamma_\tau)\coloneqq\tau^{-1}\int_0^\tau\ev{A}{\psi_t}\dd{t},
\end{equation}
where $A$ is an arbitrary time-independent operator. Unlike observable $\Phi$, which is only contributed by jumps, observable $\Lambda$ is evaluated over the entire evolution of the system's pure state.

\sectionprl{Relevant thermodynamic quantities}We discuss two quantities that play key roles in constraining the precision of observables. The first is the irreversible entropy production, which quantifies the degree of irreversibility of thermodynamic processes \cite{Landi.2021.RMP}. It is defined as the sum of the entropy changes of the system and environment as follows:
\begin{equation}
\Sigma_\tau\coloneq\Delta S_{\rm sys}+\Delta S_{\rm env},
\end{equation}
where $\Delta S_{\rm sys}$ and $\Delta S_{\rm env}$ are given by \cite{Horowitz.2013.NJP}
\begin{align}
\Delta S_{\rm sys}&\coloneq\tr{\varrho_0\ln\varrho_0}-\tr{\varrho_\tau\ln\varrho_\tau},\\
\Delta S_{\rm env}&\coloneq\int_0^\tau\sum_{k=1}^K\tr{L_k\varrho_tL_k^\dagger}\Delta s_k\dd{t}.
\end{align}
It can be shown that $\Sigma_\tau$ is always nonnegative, $\Sigma_\tau\ge 0$, which corresponds to the second law of thermodynamics.

The second key quantity is quantum dynamical activity \cite{Shiraishi.2018.PRL,Hasegawa.2020.PRL,Vu.2021.PRL}, which is quantified by the average number of jumps during period $\tau$ as follows:
\begin{equation}
\mca{A}_\tau\coloneq\sumint Np(\Gamma_\tau)\dd{\Gamma_\tau}=\ev{N}.
\end{equation}
This can be explicitly calculated as $\mca{A}_\tau=\int_0^\tau\sum_k\tr{L_k\varrho_tL_k^\dagger}\dd{t}$. Whereas entropy production is a dissipative term and measures the degree of time-reversal symmetry breaking, dynamical activity is a frenetic term that reflects the strength of thermalization of a system. These quantities complementarily characterize nonequilibrium phenomena \cite{Maes.2020.PR}.

\sectionprl{Main results}Under the aforementioned setup, we explain our main results, whose proof sketch is presented at the end of the paper (see also the Supplemental Material (SM) \cite{Supp.PhysRev} for details of the derivation). We consider a general situation in which the system is initially in an {\em arbitrary} state and aim to develop lower bounds on the relative fluctuation of observables in terms of relevant thermodynamic quantities. First, we consider an arbitrary current $J$. Using the classical Cram{\'e}r-Rao inequality \cite{Bos.2007,Hasegawa.2019.PRE}, we prove that its relative fluctuation is bounded by the entropy production and a quantum term as
\begin{equation}\label{eq:main.result.1}
\frac{\Var{J}}{\ev{J}^2}\ge\frac{2(1+\tilde{\delta} J)^2}{\Sigma_\tau+2\mca{Q}_1},
\end{equation}
where $\mca{Q}_1$ is the quantum contribution [cf.~Eq.~\eqref{eq:Q1.def}], $\tilde{\delta} J\coloneqq\ev{J}_*/\ev{J}$, and $\ev{J}_*$ is the average of the current in perturbative dynamics [cf.~Eq.~\eqref{eq:J.def}].
Here, the perturbative dynamics is obtained by modifying the original Hamiltonian and jump operators with a parameter $\theta$ as in Eq.~\eqref{eq:modif.dyn.TUR}.
Equation \eqref{eq:main.result.1} is the first main result, which is valid for arbitrary operational times and initial states as long as the local detailed balance is satisfied. 
In the absence of quantum coherence, we have $\mca{Q}_1=0$; thus, $\mca{Q}_1$ is identified as the contribution from the coherent dynamics. The inequality \eqref{eq:main.result.1} quantitatively implies that the relative fluctuation of the currents is lower bounded by both the irreversible entropy production and quantum coherence. In the classical limit (e.g., when $H=\mbb{0}$ and $L_k=\sqrt{\gamma_{mn}}\dyad{m}{n}$ with a transition rate $\gamma_{mn}>0$), it can be calculated that $\tilde{\delta} J=\delta J$ and $\mca{Q}_1=0$. Therefore, the relation \eqref{eq:main.result.1} recovers the classical TUR \cite{Liu.2020.PRL} and can be regarded as a quantum generalization of the TUR.

Next, we deal with an arbitrary generic counting observable $\Phi$ and static observable $\Lambda$.
Employing the same technique, we obtain the following bound on the precision of these observables:
\begin{equation}
\frac{\Var{O}}{\ev{O}^2}\ge\frac{(1+\delta O)^2}{\mca{A}_\tau+\mca{Q}_2}~\text{for}~O\in\qty{\Phi,\Lambda}.\label{eq:main.result.2}
\end{equation}
Here, $\mca{Q}_2$, which vanishes in the absence of quantum coherence, is identified as a coherence term \cite{Supp.PhysRev}. 
Equation \eqref{eq:main.result.2} is the second main result, implying that the precision of observables is constrained not only by dynamical activity but also by quantum coherence.
Notably, the local detailed balance is not required to obtain this result; thus, it is valid for general dynamics.
Moreover, it holds for arbitrary operational times and initial states.
In the classical limit, the coherence term $\mca{Q}_2$ equals zero, and the relation \eqref{eq:main.result.2} is reduced to the classical KUR \cite{Terlizzi.2019.JPA}.
Thus, it can be considered a quantum generalization of the KUR for counting observables.

Finally, we examine the FPT of an arbitrary counting observable $\Phi$ that can be measured in a quantum jump process, such as an optical process. For each stochastic realization, let $\tau$ be the first time at which the counting observable reaches a finite threshold value $\Phi_{\rm thr}$, that is, $\tau=\inf\qty{t\,|\,\Phi(\Gamma_t)\ge\Phi_{\rm thr}}$.
Evidently, the stopping time $\tau$ is a stochastic variable.
Assuming that the mean and variance of $\tau$ are finite, we obtain the following bound on the relative fluctuation of the FPT:
\begin{equation}\label{eq:main.result.3}
\frac{\Var{\tau}}{\ev{\tau}^2}\ge\frac{1}{\ev{N}_\tau+\mca{Q}_3}.
\end{equation}
Here, $\ev{N}_\tau$ is the average number of jumps evaluated at the stopping time, and $\mca{Q}_3$ is a quantum term that vanishes in the absence of coherence \cite{Supp.PhysRev}.
Equation \eqref{eq:main.result.3} is our third main result and implies that the precision of the FPT is constrained by both the average number of jumps up to that time and quantum coherence. It can also be regarded as a quantum generalization of the classical KUR for the FPT obtained in Ref.~\cite{Hiura.2021.PRE}.

The coherence terms $\mca{Q}_i~(i=1,2,3)$ are on par with the thermodynamic quantities in constraining the precision of observables.
They provide information for determining regions where coherence suppresses or enhances the relative fluctuation of observables.
Specifically, when $\mca{Q}_i$ is negative, it can be concluded that coherence tends to enhance the relative fluctuation of observables, whereas its positivity potentially indicates a violation of the original TUR and KUR.
In addition, these coherence terms are genuine contributions from the coherent dynamics beyond the density matrix of the system.
This is in good agreement with recent studies showing that quantum coherence in the density matrix cannot be directly related to the breaking of the TUR \cite{Kalaee.2021.PRE}.
In the SM \cite{Supp.PhysRev}, we provide simple upper bounds for the terms $\mca{Q}_1$ and $\mca{Q}_2$ in the long-time regime, which can be computed using only the Hamiltonian, jump operators, and density matrix.

We discuss the differences between the results obtained here and the related quantum bounds.
References \cite{Carollo.2019.PRL,Hasegawa.2020.PRL} derived quantum KURs for counting observables; however, they are only applicable to the steady-state and long-time limit $\tau\to\infty$, whereas our results are valid for arbitrary initial states and operational time $\tau$.
A quantum TUR for {\em instantaneous} currents in a nonequilibrium steady-state system was obtained in Ref.~\cite{Guarnieri.2019.PRR}. By contrast, our bounds apply to time-integrated observables.
In Ref.~\cite{Hasegawa.2020.PRL}, Hasegawa derived a quantum TUR for time-integrated currents [see Eq.~(14) therein]; nevertheless, Hasegawa's bound cannot be applied to quantum systems that involve quantum coherence.
In Refs.~\cite{Hasegawa.2021.PRL,Hasegawa.2021.PRL2}, Hasegawa also obtained two finite-time bounds that are applicable to arbitrary initial states in general open quantum systems; however, these two bounds are neither directly related to entropy production nor to dynamical activity, and decay exponentially in the long-time regime.
On the other hand, our bound [Eq.~\eqref{eq:main.result.1}] can characterize the interplay between dissipation and coherence in the suppression of current fluctuations, and simultaneously it reduces to the original TUR [Eq.~\eqref{eq:org.TUR}] in the classical limit.
Regarding the FPT, a quantum KUR was derived for quantum jump processes that stopped after a fixed number of jumps \cite{Hasegawa.2021.arxiv.QTURFPT}.
This type of process is a particular case of the general first passage process considered in this paper.

\sectionprl{Sufficient conditions for the classical uncertainty relations to survive}Here we reveal the deterministic effect of quantum coherence on the precision of observables, and thus determining sufficient conditions for the survival of the classical relations in the quantum regime.
We consider a {\it generic} case in which the Hamiltonian is nondegenerate and the jump operators characterize transitions between different energy eigenstates with the same energy change.
In particular, they satisfy $[L_k,H]=\omega_kL_k$, where $\omega_k=-\omega_{k'}$ denotes the energy change associated with the jump.
In this case, we prove that the coherence terms are always negative, that is, $\mca{Q}_i\le 0$ for all $i=1,2,3$ (see the SM \cite{Supp.PhysRev}).
This implies that quantum coherence has no advantage in suppressing the relative fluctuation of both the dynamical observables and the FPT, indicating the richness of the intrinsic features of thermal processes \cite{Shiraishi.2019.PRL,Abiuso.2020.E,Vu.2021.PRL2}. Notably, Eqs.~\eqref{eq:main.result.2} and \eqref{eq:main.result.3} recover the classical KURs is this generic case; moreover, Eq.~\eqref{eq:main.result.1} results exactly in the classical TUR when the system is nonresonant, thus implying that the classical relations survive under these conditions. The details of this consequence are summarized in Table \ref{table:summa}.

\begin{figure}[t]
\centering
\includegraphics[width=1.0\linewidth]{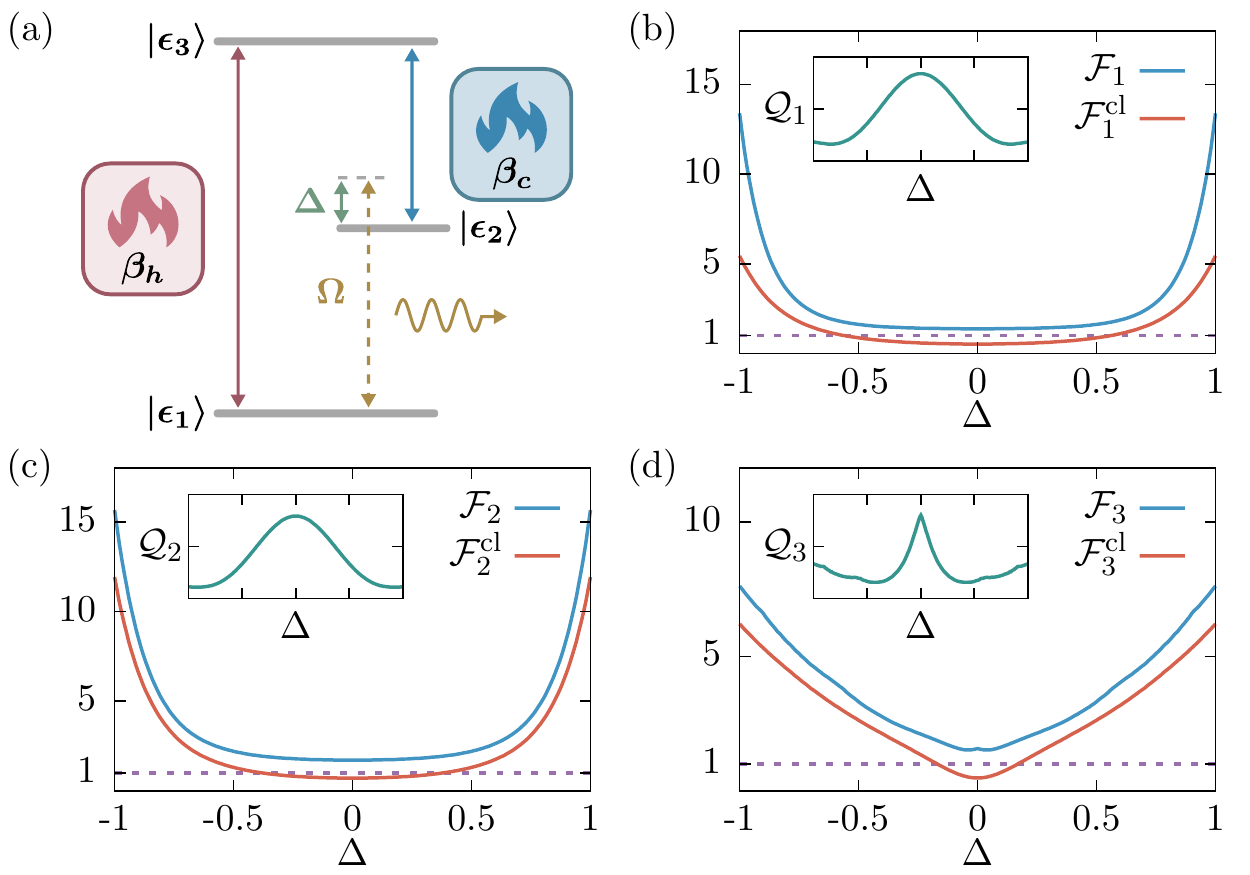}
\protect\caption{(a) Schematic of the maser. The numerical verification of the bounds in (b) Eq.~\eqref{eq:main.result.1} and (c) Eq.~\eqref{eq:main.result.2} for current $J$ and in (d) Eq.~\eqref{eq:main.result.3} for the FPT with a threshold of $J_{\rm thr}=1$. The quality factors are depicted as solid lines. The insets show the variation in the coherence terms $\mca{Q}_i$ as $\Delta$ is changed. Other parameters are fixed as $\gamma_h=0.1$, $\gamma_c=2.0$, $n_h=5.0$, $n_c=0.01$, $\Omega=0.15$, and $\tau=5$.}\label{fig:result1}
\end{figure}

\sectionprl{Example}We illustrate the main results in a three-level maser, which interacts with a classical electric field and a hot and a cold heat bath [see Fig.~\ref{fig:result1}(a)].
The maser can operate as a heat engine or refrigerator, depending on the parameters.
To guarantee the validity of the local master equation, we exclusively consider the weak driving field \cite{Geva.1996.JCP}.
In a rotating frame \cite{Boukobza.2007.PRL,Kalaee.2021.PRE}, the dynamics of the density matrix is governed by the Hamiltonian $H=-\Delta\sigma_{22}+\Omega(\sigma_{12}+\sigma_{21})$ and the jump operators $L_1=\sqrt{\gamma_hn_h}\sigma_{31}$, $L_{1'}=\sqrt{\gamma_h(n_h+1)}\sigma_{13}$, $L_2=\sqrt{\gamma_cn_c}\sigma_{32}$, and $L_{2'}=\sqrt{\gamma_c(n_c+1)}\sigma_{23}$.
Here, $\Delta$ is a detuning parameter, $\Omega$ is the coupling strength of the driving field, $\sigma_{ij}\coloneqq\dyad{\epsilon_i}{\epsilon_j}$, and $\gamma_x$ and $n_x$ are the decay rate and the thermal occupation number for $x\in\qty{c,h}$, respectively.

We consider a current $J$ with $\vec{w}=[1,-1,-1,1]^\top$, which is proportional to the net number of cycles, and the FPT of the current with a threshold of $J_{\rm thr}=1$.
The quality factor of each bound is defined as the relative fluctuation divided by the lower-bound term, which should be greater than or equal to $1$.
Let $\mca{F}_i~(i=1,2,3)$ be the quality factors associated with the derived bounds in Eqs.~\eqref{eq:main.result.1}, \eqref{eq:main.result.2}, and \eqref{eq:main.result.3}, respectively, and $\mca{F}_i^{\rm cl}$ are the factors of the corresponding classical TUR or KUR.
We vary $\Delta$ while fixing the other parameters. The initial state is set to a pure state $\varrho_0=\dyad{\epsilon_2}$. The quality factors of the bounds are numerically evaluated using $10^6$ trajectories and are plotted in Figs.~\ref{fig:result1}(b)--\ref{fig:result1}(d).
As shown, $\mca{F}_i$ is always greater than $1$, which numerically validates the derived bounds.
In contrast, the classical bounds are significantly violated for $|\Delta|\le\Omega\,(=0.15)$.
Although we focus here on the transient dynamics, the same violation was also observed in steady-state dynamics \cite{Kalaee.2021.PRE,Menczel.2021.JPA}.
In the region where the classical bounds are invalid, the coherence terms become relatively large compared to the thermodynamic quantities, showing a quantum advantage in enhancing the precision of observables.

\sectionprl{Sketch of proof}We consider an auxiliary dynamics, which is obtained by perturbing the original dynamics [Eq.~\eqref{eq:Lindblad.dyn}] by a parameter $\theta$; when $\theta=0$, the auxiliary dynamics is reduced to the original.
According to the classical Cram{\'e}r-Rao inequality, we have
\begin{equation}\label{eq:CR.ineq}
\frac{\Var{\Phi}}{(\partial_\theta\ev{\Phi}_\theta|_{\theta=0})^2}\ge\frac{1}{\mca{I}(0)},
\end{equation}
where the Fisher information is given by $\mca{I}(0)=-\ev{\partial_\theta^2\ln p_\theta(\Gamma_\tau)}|_{\theta=0}$ and the subscript $\theta$ is associated with auxiliary dynamics.
To derive Eq.~\eqref{eq:main.result.1}, we consider the auxiliary dynamics with the Hamiltonian and jump operators modified as
\begin{equation}\label{eq:modif.dyn.TUR}
H_\theta(t)=(1+\theta)H,~L_{k,\theta}(t)=\sqrt{1+\ell_k(t)\theta}L_k,
\end{equation}
where the coefficient $\ell_k(t)$ is given by
\begin{equation}
\ell_k(t)=\frac{\tr{L_{k}\varrho_tL_{k}^\dagger}-\tr{L_{k'}\varrho_tL_{k'}^\dagger}}{\tr{L_{k}\varrho_tL_{k}^\dagger}+\tr{L_{k'}\varrho_tL_{k'}^\dagger}}.
\end{equation}
With this modification, the Fisher information can be upper bounded as $\mca{I}(0)\le\Sigma_\tau/2+\mca{Q}_1$, where
\begin{align}
\mca{Q}_1&\coloneqq -\ev{\partial_\theta^2\ln\qty|\ket{\Psi_\theta(\Gamma_\tau)}|^2}_{\theta=0},\label{eq:Q1.def}\\
\ket{\Psi_\theta(\Gamma_\tau)}&\coloneqq U_{\theta}(\tau,t_N)\prod_{j=1}^{N}L_{k_{j}}U_{\theta}(t_{j},t_{j-1})\ket{n}.
\end{align}
In addition, $\partial_\theta\ev{J}_\theta|_{\theta=0}=\ev{J}+\ev{J}_*$, where
\begin{equation}\label{eq:J.def}
\ev{J}_*\coloneqq\int_0^\tau\sum_kw_k\tr{L_k\phi_tL_k^\dagger}\dd{t}
\end{equation}
and the traceless operator $\phi_t$ evolves according to the equation $\dot\phi_t=\mca{L}(\varrho_t+\phi_t)+\sum_k[\ell_k(t)-1]\mca{D}[L_k]\varrho_t$ with the initial condition $\phi_0=\mbb{0}$.
Likewise, Eqs.~\eqref{eq:main.result.2} and \eqref{eq:main.result.3} are derived using the following auxiliary dynamics:
\begin{equation}
H_\theta(t)=(1+\theta)H,~L_{k,\theta}(t)=\sqrt{1+\theta}L_k.
\end{equation}
In this case, we obtain $\mca{I}(0)=\mca{A}_\tau+\mca{Q}_2$ for counting and static observables and $\mca{I}(0)=\ev{N}_\tau+\mca{Q}_3$ for the FPT observable.
Here, $\mca{Q}_2$ and $\mca{Q}_3$ are defined analogously as in Eq.~\eqref{eq:Q1.def} with the corresponding time propagation operator $U_\theta$.

\sectionprl{Summary}In this paper, we derived fundamental bounds on the precision of dynamical and time observables for Markovian dynamics. These bounds indicate that quantum coherence plays a key role in constraining the relative fluctuation of the observables. Moreover, they provide insights into the precision of thermal machines, such as heat engines and quantum clocks. Restricting to the generic class of dissipative processes, we found that quantum coherence tends to enhance the relative fluctuation of the observables. Because thermal processes are relevant in heat engines, our bounds can be applied to obtain useful trade-off relations between the power and efficiency.

\begin{acknowledgments}
We thank Y. Hasegawa, T. Kuwahara, and M. Kewming for fruitful discussions.
We also acknowledge the anonymous referees for invaluable comments on the manuscript.
This work was supported by Grants-in-Aid for Scientific Research (JP19H05603 and JP19H05791).
\end{acknowledgments}


\begin{thebibliography}{95}%
\makeatletter
\providecommand \@ifxundefined [1]{%
 \@ifx{#1\undefined}
}%
\providecommand \@ifnum [1]{%
 \ifnum #1\expandafter \@firstoftwo
 \else \expandafter \@secondoftwo
 \fi
}%
\providecommand \@ifx [1]{%
 \ifx #1\expandafter \@firstoftwo
 \else \expandafter \@secondoftwo
 \fi
}%
\providecommand \natexlab [1]{#1}%
\providecommand \enquote  [1]{``#1''}%
\providecommand \bibnamefont  [1]{#1}%
\providecommand \bibfnamefont [1]{#1}%
\providecommand \citenamefont [1]{#1}%
\providecommand \href@noop [0]{\@secondoftwo}%
\providecommand \href [0]{\begingroup \@sanitize@url \@href}%
\providecommand \@href[1]{\@@startlink{#1}\@@href}%
\providecommand \@@href[1]{\endgroup#1\@@endlink}%
\providecommand \@sanitize@url [0]{\catcode `\\12\catcode `\$12\catcode
  `\&12\catcode `\#12\catcode `\^12\catcode `\_12\catcode `\%12\relax}%
\providecommand \@@startlink[1]{}%
\providecommand \@@endlink[0]{}%
\providecommand \url  [0]{\begingroup\@sanitize@url \@url }%
\providecommand \@url [1]{\endgroup\@href {#1}{\urlprefix }}%
\providecommand \urlprefix  [0]{URL }%
\providecommand \Eprint [0]{\href }%
\providecommand \doibase [0]{https://doi.org/}%
\providecommand \selectlanguage [0]{\@gobble}%
\providecommand \bibinfo  [0]{\@secondoftwo}%
\providecommand \bibfield  [0]{\@secondoftwo}%
\providecommand \translation [1]{[#1]}%
\providecommand \BibitemOpen [0]{}%
\providecommand \bibitemStop [0]{}%
\providecommand \bibitemNoStop [0]{.\EOS\space}%
\providecommand \EOS [0]{\spacefactor3000\relax}%
\providecommand \BibitemShut  [1]{\csname bibitem#1\endcsname}%
\let\auto@bib@innerbib\@empty
\bibitem [{\citenamefont {Evans}\ \emph {et~al.}(1993)\citenamefont {Evans},
  \citenamefont {Cohen},\ and\ \citenamefont {Morriss}}]{Evans.1993.PRL}%
  \BibitemOpen
  \bibfield  {author} {\bibinfo {author} {\bibfnamefont {D.~J.}\ \bibnamefont
  {Evans}}, \bibinfo {author} {\bibfnamefont {E.~G.~D.}\ \bibnamefont
  {Cohen}},\ and\ \bibinfo {author} {\bibfnamefont {G.~P.}\ \bibnamefont
  {Morriss}},\ }\bibfield  {title} {\bibinfo {title} {{Probability of second
  law violations in shearing steady states}},\ }\href
  {https://doi.org/10.1103/PhysRevLett.71.2401} {\bibfield  {journal} {\bibinfo
   {journal} {Phys. Rev. Lett.}\ }\textbf {\bibinfo {volume} {71}},\ \bibinfo
  {pages} {2401} (\bibinfo {year} {1993})}\BibitemShut {NoStop}%
\bibitem [{\citenamefont {Gallavotti}\ and\ \citenamefont
  {Cohen}(1995)}]{Gallavotti.1995.PRL}%
  \BibitemOpen
  \bibfield  {author} {\bibinfo {author} {\bibfnamefont {G.}~\bibnamefont
  {Gallavotti}}\ and\ \bibinfo {author} {\bibfnamefont {E.~G.~D.}\ \bibnamefont
  {Cohen}},\ }\bibfield  {title} {\bibinfo {title} {{Dynamical ensembles in
  nonequilibrium statistical mechanics}},\ }\href
  {https://doi.org/10.1103/PhysRevLett.74.2694} {\bibfield  {journal} {\bibinfo
   {journal} {Phys. Rev. Lett.}\ }\textbf {\bibinfo {volume} {74}},\ \bibinfo
  {pages} {2694} (\bibinfo {year} {1995})}\BibitemShut {NoStop}%
\bibitem [{\citenamefont {Crooks}(1999)}]{Crooks.1999.PRE}%
  \BibitemOpen
  \bibfield  {author} {\bibinfo {author} {\bibfnamefont {G.~E.}\ \bibnamefont
  {Crooks}},\ }\bibfield  {title} {\bibinfo {title} {{Entropy production
  fluctuation theorem and the nonequilibrium work relation for free energy
  differences}},\ }\href {https://doi.org/10.1103/PhysRevE.60.2721} {\bibfield
  {journal} {\bibinfo  {journal} {Phys. Rev. E}\ }\textbf {\bibinfo {volume}
  {60}},\ \bibinfo {pages} {2721} (\bibinfo {year} {1999})}\BibitemShut
  {NoStop}%
\bibitem [{\citenamefont {Jarzynski}(2000)}]{Jarzynski.2000.JSP}%
  \BibitemOpen
  \bibfield  {author} {\bibinfo {author} {\bibfnamefont {C.}~\bibnamefont
  {Jarzynski}},\ }\bibfield  {title} {\bibinfo {title} {{Hamiltonian derivation
  of a detailed fluctuation theorem}},\ }\href
  {https://doi.org/10.1023/a:1018670721277} {\bibfield  {journal} {\bibinfo
  {journal} {J. Stat. Phys.}\ }\textbf {\bibinfo {volume} {98}},\ \bibinfo
  {pages} {77} (\bibinfo {year} {2000})}\BibitemShut {NoStop}%
\bibitem [{\citenamefont {Esposito}\ \emph {et~al.}(2009)\citenamefont
  {Esposito}, \citenamefont {Harbola},\ and\ \citenamefont
  {Mukamel}}]{Esposito.2009.RMP}%
  \BibitemOpen
  \bibfield  {author} {\bibinfo {author} {\bibfnamefont {M.}~\bibnamefont
  {Esposito}}, \bibinfo {author} {\bibfnamefont {U.}~\bibnamefont {Harbola}},\
  and\ \bibinfo {author} {\bibfnamefont {S.}~\bibnamefont {Mukamel}},\
  }\bibfield  {title} {\bibinfo {title} {{Nonequilibrium fluctuations,
  fluctuation theorems, and counting statistics in quantum systems}},\ }\href
  {https://doi.org/10.1103/RevModPhys.81.1665} {\bibfield  {journal} {\bibinfo
  {journal} {Rev. Mod. Phys.}\ }\textbf {\bibinfo {volume} {81}},\ \bibinfo
  {pages} {1665} (\bibinfo {year} {2009})}\BibitemShut {NoStop}%
\bibitem [{\citenamefont {Campisi}\ \emph {et~al.}(2011)\citenamefont
  {Campisi}, \citenamefont {H\"anggi},\ and\ \citenamefont
  {Talkner}}]{Campisi.2011.RMP}%
  \BibitemOpen
  \bibfield  {author} {\bibinfo {author} {\bibfnamefont {M.}~\bibnamefont
  {Campisi}}, \bibinfo {author} {\bibfnamefont {P.}~\bibnamefont {H\"anggi}},\
  and\ \bibinfo {author} {\bibfnamefont {P.}~\bibnamefont {Talkner}},\
  }\bibfield  {title} {\bibinfo {title} {{Colloquium: Quantum fluctuation
  relations: Foundations and applications}},\ }\href
  {https://doi.org/10.1103/RevModPhys.83.771} {\bibfield  {journal} {\bibinfo
  {journal} {Rev. Mod. Phys.}\ }\textbf {\bibinfo {volume} {83}},\ \bibinfo
  {pages} {771} (\bibinfo {year} {2011})}\BibitemShut {NoStop}%
\bibitem [{\citenamefont {Barato}\ and\ \citenamefont
  {Seifert}(2015)}]{Barato.2015.PRL}%
  \BibitemOpen
  \bibfield  {author} {\bibinfo {author} {\bibfnamefont {A.~C.}\ \bibnamefont
  {Barato}}\ and\ \bibinfo {author} {\bibfnamefont {U.}~\bibnamefont
  {Seifert}},\ }\bibfield  {title} {\bibinfo {title} {{Thermodynamic
  uncertainty relation for biomolecular processes}},\ }\href
  {https://doi.org/10.1103/PhysRevLett.114.158101} {\bibfield  {journal}
  {\bibinfo  {journal} {Phys. Rev. Lett.}\ }\textbf {\bibinfo {volume} {114}},\
  \bibinfo {pages} {158101} (\bibinfo {year} {2015})}\BibitemShut {NoStop}%
\bibitem [{\citenamefont {Gingrich}\ \emph {et~al.}(2016)\citenamefont
  {Gingrich}, \citenamefont {Horowitz}, \citenamefont {Perunov},\ and\
  \citenamefont {England}}]{Gingrich.2016.PRL}%
  \BibitemOpen
  \bibfield  {author} {\bibinfo {author} {\bibfnamefont {T.~R.}\ \bibnamefont
  {Gingrich}}, \bibinfo {author} {\bibfnamefont {J.~M.}\ \bibnamefont
  {Horowitz}}, \bibinfo {author} {\bibfnamefont {N.}~\bibnamefont {Perunov}},\
  and\ \bibinfo {author} {\bibfnamefont {J.~L.}\ \bibnamefont {England}},\
  }\bibfield  {title} {\bibinfo {title} {{Dissipation bounds all steady-state
  current fluctuations}},\ }\href
  {https://doi.org/10.1103/PhysRevLett.116.120601} {\bibfield  {journal}
  {\bibinfo  {journal} {Phys. Rev. Lett.}\ }\textbf {\bibinfo {volume} {116}},\
  \bibinfo {pages} {120601} (\bibinfo {year} {2016})}\BibitemShut {NoStop}%
\bibitem [{\citenamefont {Horowitz}\ and\ \citenamefont
  {Gingrich}(2017)}]{Horowitz.2017.PRE}%
  \BibitemOpen
  \bibfield  {author} {\bibinfo {author} {\bibfnamefont {J.~M.}\ \bibnamefont
  {Horowitz}}\ and\ \bibinfo {author} {\bibfnamefont {T.~R.}\ \bibnamefont
  {Gingrich}},\ }\bibfield  {title} {\bibinfo {title} {{Proof of the
  finite-time thermodynamic uncertainty relation for steady-state currents}},\
  }\href {https://doi.org/10.1103/PhysRevE.96.020103} {\bibfield  {journal}
  {\bibinfo  {journal} {Phys. Rev. E}\ }\textbf {\bibinfo {volume} {96}},\
  \bibinfo {pages} {020103(R)} (\bibinfo {year} {2017})}\BibitemShut {NoStop}%
\bibitem [{\citenamefont {Dechant}\ and\ \citenamefont
  {Sasa}(2018)}]{Dechant.2018.JSM}%
  \BibitemOpen
  \bibfield  {author} {\bibinfo {author} {\bibfnamefont {A.}~\bibnamefont
  {Dechant}}\ and\ \bibinfo {author} {\bibfnamefont {S.-i.}\ \bibnamefont
  {Sasa}},\ }\bibfield  {title} {\bibinfo {title} {{Current fluctuations and
  transport efficiency for general Langevin systems}},\ }\href
  {https://doi.org/10.1088/1742-5468/aac91a} {\bibfield  {journal} {\bibinfo
  {journal} {J. Stat. Mech.: Theory Exp.}\ }\textbf {\bibinfo {volume}
  {2018}},\ \bibinfo {pages} {063209}}\BibitemShut {NoStop}%
\bibitem [{\citenamefont {Liu}\ \emph {et~al.}(2020)\citenamefont {Liu},
  \citenamefont {Gong},\ and\ \citenamefont {Ueda}}]{Liu.2020.PRL}%
  \BibitemOpen
  \bibfield  {author} {\bibinfo {author} {\bibfnamefont {K.}~\bibnamefont
  {Liu}}, \bibinfo {author} {\bibfnamefont {Z.}~\bibnamefont {Gong}},\ and\
  \bibinfo {author} {\bibfnamefont {M.}~\bibnamefont {Ueda}},\ }\bibfield
  {title} {\bibinfo {title} {{Thermodynamic uncertainty relation for arbitrary
  initial states}},\ }\href {https://doi.org/10.1103/PhysRevLett.125.140602}
  {\bibfield  {journal} {\bibinfo  {journal} {Phys. Rev. Lett.}\ }\textbf
  {\bibinfo {volume} {125}},\ \bibinfo {pages} {140602} (\bibinfo {year}
  {2020})}\BibitemShut {NoStop}%
\bibitem [{\citenamefont {Garrahan}(2017)}]{Garrahan.2017.PRE}%
  \BibitemOpen
  \bibfield  {author} {\bibinfo {author} {\bibfnamefont {J.~P.}\ \bibnamefont
  {Garrahan}},\ }\bibfield  {title} {\bibinfo {title} {{Simple bounds on
  fluctuations and uncertainty relations for first-passage times of counting
  observables}},\ }\href {https://doi.org/10.1103/PhysRevE.95.032134}
  {\bibfield  {journal} {\bibinfo  {journal} {Phys. Rev. E}\ }\textbf {\bibinfo
  {volume} {95}},\ \bibinfo {pages} {032134} (\bibinfo {year}
  {2017})}\BibitemShut {NoStop}%
\bibitem [{\citenamefont {Terlizzi}\ and\ \citenamefont
  {Baiesi}(2019)}]{Terlizzi.2019.JPA}%
  \BibitemOpen
  \bibfield  {author} {\bibinfo {author} {\bibfnamefont {I.~D.}\ \bibnamefont
  {Terlizzi}}\ and\ \bibinfo {author} {\bibfnamefont {M.}~\bibnamefont
  {Baiesi}},\ }\bibfield  {title} {\bibinfo {title} {{Kinetic uncertainty
  relation}},\ }\href {https://doi.org/10.1088/1751-8121/aaee34} {\bibfield
  {journal} {\bibinfo  {journal} {J. Phys. A}\ }\textbf {\bibinfo {volume}
  {52}},\ \bibinfo {pages} {02LT03} (\bibinfo {year} {2019})}\BibitemShut
  {NoStop}%
\bibitem [{\citenamefont {Pietzonka}\ and\ \citenamefont
  {Seifert}(2018)}]{Pietzonka.2018.PRL}%
  \BibitemOpen
  \bibfield  {author} {\bibinfo {author} {\bibfnamefont {P.}~\bibnamefont
  {Pietzonka}}\ and\ \bibinfo {author} {\bibfnamefont {U.}~\bibnamefont
  {Seifert}},\ }\bibfield  {title} {\bibinfo {title} {{Universal trade-off
  between power, efficiency, and constancy in steady-state heat engines}},\
  }\href {https://doi.org/10.1103/PhysRevLett.120.190602} {\bibfield  {journal}
  {\bibinfo  {journal} {Phys. Rev. Lett.}\ }\textbf {\bibinfo {volume} {120}},\
  \bibinfo {pages} {190602} (\bibinfo {year} {2018})}\BibitemShut {NoStop}%
\bibitem [{\citenamefont {Shiraishi}\ \emph {et~al.}(2016)\citenamefont
  {Shiraishi}, \citenamefont {Saito},\ and\ \citenamefont
  {Tasaki}}]{Shiraishi.2016.PRL}%
  \BibitemOpen
  \bibfield  {author} {\bibinfo {author} {\bibfnamefont {N.}~\bibnamefont
  {Shiraishi}}, \bibinfo {author} {\bibfnamefont {K.}~\bibnamefont {Saito}},\
  and\ \bibinfo {author} {\bibfnamefont {H.}~\bibnamefont {Tasaki}},\
  }\bibfield  {title} {\bibinfo {title} {{Universal trade-off relation between
  power and efficiency for heat engines}},\ }\href
  {https://doi.org/10.1103/PhysRevLett.117.190601} {\bibfield  {journal}
  {\bibinfo  {journal} {Phys. Rev. Lett.}\ }\textbf {\bibinfo {volume} {117}},\
  \bibinfo {pages} {190601} (\bibinfo {year} {2016})}\BibitemShut {NoStop}%
\bibitem [{\citenamefont {Li}\ \emph {et~al.}(2019)\citenamefont {Li},
  \citenamefont {Horowitz}, \citenamefont {Gingrich},\ and\ \citenamefont
  {Fakhri}}]{Li.2019.NC}%
  \BibitemOpen
  \bibfield  {author} {\bibinfo {author} {\bibfnamefont {J.}~\bibnamefont
  {Li}}, \bibinfo {author} {\bibfnamefont {J.~M.}\ \bibnamefont {Horowitz}},
  \bibinfo {author} {\bibfnamefont {T.~R.}\ \bibnamefont {Gingrich}},\ and\
  \bibinfo {author} {\bibfnamefont {N.}~\bibnamefont {Fakhri}},\ }\bibfield
  {title} {\bibinfo {title} {{Quantifying dissipation using fluctuating
  currents}},\ }\href {https://doi.org/10.1038/s41467-019-09631-x} {\bibfield
  {journal} {\bibinfo  {journal} {Nat. Commun.}\ }\textbf {\bibinfo {volume}
  {10}},\ \bibinfo {pages} {1666} (\bibinfo {year} {2019})}\BibitemShut
  {NoStop}%
\bibitem [{\citenamefont {Manikandan}\ \emph {et~al.}(2020)\citenamefont
  {Manikandan}, \citenamefont {Gupta},\ and\ \citenamefont
  {Krishnamurthy}}]{Manikandan.2020.PRL}%
  \BibitemOpen
  \bibfield  {author} {\bibinfo {author} {\bibfnamefont {S.~K.}\ \bibnamefont
  {Manikandan}}, \bibinfo {author} {\bibfnamefont {D.}~\bibnamefont {Gupta}},\
  and\ \bibinfo {author} {\bibfnamefont {S.}~\bibnamefont {Krishnamurthy}},\
  }\bibfield  {title} {\bibinfo {title} {{Inferring entropy production from
  short experiments}},\ }\href {https://doi.org/10.1103/PhysRevLett.124.120603}
  {\bibfield  {journal} {\bibinfo  {journal} {Phys. Rev. Lett.}\ }\textbf
  {\bibinfo {volume} {124}},\ \bibinfo {pages} {120603} (\bibinfo {year}
  {2020})}\BibitemShut {NoStop}%
\bibitem [{\citenamefont {Van~Vu}\ \emph {et~al.}(2020)\citenamefont {Van~Vu},
  \citenamefont {Vo},\ and\ \citenamefont {Hasegawa}}]{Vu.2020.PRE}%
  \BibitemOpen
  \bibfield  {author} {\bibinfo {author} {\bibfnamefont {T.}~\bibnamefont
  {Van~Vu}}, \bibinfo {author} {\bibfnamefont {V.~T.}\ \bibnamefont {Vo}},\
  and\ \bibinfo {author} {\bibfnamefont {Y.}~\bibnamefont {Hasegawa}},\
  }\bibfield  {title} {\bibinfo {title} {{Entropy production estimation with
  optimal current}},\ }\href {https://doi.org/10.1103/PhysRevE.101.042138}
  {\bibfield  {journal} {\bibinfo  {journal} {Phys. Rev. E}\ }\textbf {\bibinfo
  {volume} {101}},\ \bibinfo {pages} {042138} (\bibinfo {year}
  {2020})}\BibitemShut {NoStop}%
\bibitem [{\citenamefont {Otsubo}\ \emph {et~al.}(2020)\citenamefont {Otsubo},
  \citenamefont {Ito}, \citenamefont {Dechant},\ and\ \citenamefont
  {Sagawa}}]{Otsubo.2020.PRE}%
  \BibitemOpen
  \bibfield  {author} {\bibinfo {author} {\bibfnamefont {S.}~\bibnamefont
  {Otsubo}}, \bibinfo {author} {\bibfnamefont {S.}~\bibnamefont {Ito}},
  \bibinfo {author} {\bibfnamefont {A.}~\bibnamefont {Dechant}},\ and\ \bibinfo
  {author} {\bibfnamefont {T.}~\bibnamefont {Sagawa}},\ }\bibfield  {title}
  {\bibinfo {title} {{Estimating entropy production by machine learning of
  short-time fluctuating currents}},\ }\href
  {https://doi.org/10.1103/PhysRevE.101.062106} {\bibfield  {journal} {\bibinfo
   {journal} {Phys. Rev. E}\ }\textbf {\bibinfo {volume} {101}},\ \bibinfo
  {pages} {062106} (\bibinfo {year} {2020})}\BibitemShut {NoStop}%
\bibitem [{\citenamefont {Gingrich}\ and\ \citenamefont
  {Horowitz}(2017)}]{Gingrich.2017.PRL}%
  \BibitemOpen
  \bibfield  {author} {\bibinfo {author} {\bibfnamefont {T.~R.}\ \bibnamefont
  {Gingrich}}\ and\ \bibinfo {author} {\bibfnamefont {J.~M.}\ \bibnamefont
  {Horowitz}},\ }\bibfield  {title} {\bibinfo {title} {{Fundamental bounds on
  first passage time fluctuations for currents}},\ }\href
  {https://doi.org/10.1103/PhysRevLett.119.170601} {\bibfield  {journal}
  {\bibinfo  {journal} {Phys. Rev. Lett.}\ }\textbf {\bibinfo {volume} {119}},\
  \bibinfo {pages} {170601} (\bibinfo {year} {2017})}\BibitemShut {NoStop}%
\bibitem [{\citenamefont {Brandner}\ \emph {et~al.}(2018)\citenamefont
  {Brandner}, \citenamefont {Hanazato},\ and\ \citenamefont
  {Saito}}]{Brandner.2018.PRL}%
  \BibitemOpen
  \bibfield  {author} {\bibinfo {author} {\bibfnamefont {K.}~\bibnamefont
  {Brandner}}, \bibinfo {author} {\bibfnamefont {T.}~\bibnamefont {Hanazato}},\
  and\ \bibinfo {author} {\bibfnamefont {K.}~\bibnamefont {Saito}},\ }\bibfield
   {title} {\bibinfo {title} {{Thermodynamic bounds on precision in ballistic
  multiterminal transport}},\ }\href
  {https://doi.org/10.1103/PhysRevLett.120.090601} {\bibfield  {journal}
  {\bibinfo  {journal} {Phys. Rev. Lett.}\ }\textbf {\bibinfo {volume} {120}},\
  \bibinfo {pages} {090601} (\bibinfo {year} {2018})}\BibitemShut {NoStop}%
\bibitem [{\citenamefont {Agarwalla}\ and\ \citenamefont
  {Segal}(2018)}]{Agarwalla.2018.PRB}%
  \BibitemOpen
  \bibfield  {author} {\bibinfo {author} {\bibfnamefont {B.~K.}\ \bibnamefont
  {Agarwalla}}\ and\ \bibinfo {author} {\bibfnamefont {D.}~\bibnamefont
  {Segal}},\ }\bibfield  {title} {\bibinfo {title} {{Assessing the validity of
  the thermodynamic uncertainty relation in quantum systems}},\ }\href
  {https://doi.org/10.1103/PhysRevB.98.155438} {\bibfield  {journal} {\bibinfo
  {journal} {Phys. Rev. B}\ }\textbf {\bibinfo {volume} {98}},\ \bibinfo
  {pages} {155438} (\bibinfo {year} {2018})}\BibitemShut {NoStop}%
\bibitem [{\citenamefont {Hasegawa}\ and\ \citenamefont
  {Van~Vu}(2019{\natexlab{a}})}]{Hasegawa.2019.PRE}%
  \BibitemOpen
  \bibfield  {author} {\bibinfo {author} {\bibfnamefont {Y.}~\bibnamefont
  {Hasegawa}}\ and\ \bibinfo {author} {\bibfnamefont {T.}~\bibnamefont
  {Van~Vu}},\ }\bibfield  {title} {\bibinfo {title} {{Uncertainty relations in
  stochastic processes: An information inequality approach}},\ }\href
  {https://doi.org/10.1103/PhysRevE.99.062126} {\bibfield  {journal} {\bibinfo
  {journal} {Phys. Rev. E}\ }\textbf {\bibinfo {volume} {99}},\ \bibinfo
  {pages} {062126} (\bibinfo {year} {2019}{\natexlab{a}})}\BibitemShut
  {NoStop}%
\bibitem [{\citenamefont {Saryal}\ \emph {et~al.}(2019)\citenamefont {Saryal},
  \citenamefont {Friedman}, \citenamefont {Segal},\ and\ \citenamefont
  {Agarwalla}}]{Saryal.2019.PRE}%
  \BibitemOpen
  \bibfield  {author} {\bibinfo {author} {\bibfnamefont {S.}~\bibnamefont
  {Saryal}}, \bibinfo {author} {\bibfnamefont {H.~M.}\ \bibnamefont
  {Friedman}}, \bibinfo {author} {\bibfnamefont {D.}~\bibnamefont {Segal}},\
  and\ \bibinfo {author} {\bibfnamefont {B.~K.}\ \bibnamefont {Agarwalla}},\
  }\bibfield  {title} {\bibinfo {title} {{Thermodynamic uncertainty relation in
  thermal transport}},\ }\href {https://doi.org/10.1103/PhysRevE.100.042101}
  {\bibfield  {journal} {\bibinfo  {journal} {Phys. Rev. E}\ }\textbf {\bibinfo
  {volume} {100}},\ \bibinfo {pages} {042101} (\bibinfo {year}
  {2019})}\BibitemShut {NoStop}%
\bibitem [{\citenamefont {Van~Vu}\ and\ \citenamefont
  {Hasegawa}(2019)}]{Vu.2019.PRE.UnderdampedTUR}%
  \BibitemOpen
  \bibfield  {author} {\bibinfo {author} {\bibfnamefont {T.}~\bibnamefont
  {Van~Vu}}\ and\ \bibinfo {author} {\bibfnamefont {Y.}~\bibnamefont
  {Hasegawa}},\ }\bibfield  {title} {\bibinfo {title} {{Uncertainty relations
  for underdamped Langevin dynamics}},\ }\href
  {https://doi.org/10.1103/PhysRevE.100.032130} {\bibfield  {journal} {\bibinfo
   {journal} {Phys. Rev. E}\ }\textbf {\bibinfo {volume} {100}},\ \bibinfo
  {pages} {032130} (\bibinfo {year} {2019})}\BibitemShut {NoStop}%
\bibitem [{\citenamefont {Liu}\ and\ \citenamefont
  {Segal}(2019)}]{Liu.2019.PRE}%
  \BibitemOpen
  \bibfield  {author} {\bibinfo {author} {\bibfnamefont {J.}~\bibnamefont
  {Liu}}\ and\ \bibinfo {author} {\bibfnamefont {D.}~\bibnamefont {Segal}},\
  }\bibfield  {title} {\bibinfo {title} {{Thermodynamic uncertainty relation in
  quantum thermoelectric junctions}},\ }\href
  {https://doi.org/10.1103/PhysRevE.99.062141} {\bibfield  {journal} {\bibinfo
  {journal} {Phys. Rev. E}\ }\textbf {\bibinfo {volume} {99}},\ \bibinfo
  {pages} {062141} (\bibinfo {year} {2019})}\BibitemShut {NoStop}%
\bibitem [{\citenamefont {Hasegawa}\ and\ \citenamefont
  {Van~Vu}(2019{\natexlab{b}})}]{Hasegawa.2019.PRL}%
  \BibitemOpen
  \bibfield  {author} {\bibinfo {author} {\bibfnamefont {Y.}~\bibnamefont
  {Hasegawa}}\ and\ \bibinfo {author} {\bibfnamefont {T.}~\bibnamefont
  {Van~Vu}},\ }\bibfield  {title} {\bibinfo {title} {{Fluctuation theorem
  uncertainty relation}},\ }\href
  {https://doi.org/10.1103/PhysRevLett.123.110602} {\bibfield  {journal}
  {\bibinfo  {journal} {Phys. Rev. Lett.}\ }\textbf {\bibinfo {volume} {123}},\
  \bibinfo {pages} {110602} (\bibinfo {year} {2019}{\natexlab{b}})}\BibitemShut
  {NoStop}%
\bibitem [{\citenamefont {Timpanaro}\ \emph {et~al.}(2019)\citenamefont
  {Timpanaro}, \citenamefont {Guarnieri}, \citenamefont {Goold},\ and\
  \citenamefont {Landi}}]{Timpanaro.2019.PRL}%
  \BibitemOpen
  \bibfield  {author} {\bibinfo {author} {\bibfnamefont {A.~M.}\ \bibnamefont
  {Timpanaro}}, \bibinfo {author} {\bibfnamefont {G.}~\bibnamefont
  {Guarnieri}}, \bibinfo {author} {\bibfnamefont {J.}~\bibnamefont {Goold}},\
  and\ \bibinfo {author} {\bibfnamefont {G.~T.}\ \bibnamefont {Landi}},\
  }\bibfield  {title} {\bibinfo {title} {{Thermodynamic uncertainty relations
  from exchange fluctuation theorems}},\ }\href
  {https://doi.org/10.1103/PhysRevLett.123.090604} {\bibfield  {journal}
  {\bibinfo  {journal} {Phys. Rev. Lett.}\ }\textbf {\bibinfo {volume} {123}},\
  \bibinfo {pages} {090604} (\bibinfo {year} {2019})}\BibitemShut {NoStop}%
\bibitem [{\citenamefont {Horowitz}\ and\ \citenamefont
  {Gingrich}(2020)}]{Horowitz.2020.NP}%
  \BibitemOpen
  \bibfield  {author} {\bibinfo {author} {\bibfnamefont {J.~M.}\ \bibnamefont
  {Horowitz}}\ and\ \bibinfo {author} {\bibfnamefont {T.~R.}\ \bibnamefont
  {Gingrich}},\ }\bibfield  {title} {\bibinfo {title} {{Thermodynamic
  uncertainty relations constrain non-equilibrium fluctuations}},\ }\href
  {https://doi.org/10.1038/s41567-019-0702-6} {\bibfield  {journal} {\bibinfo
  {journal} {Nat. Phys.}\ }\textbf {\bibinfo {volume} {16}},\ \bibinfo {pages}
  {15} (\bibinfo {year} {2020})}\BibitemShut {NoStop}%
\bibitem [{\citenamefont {Van~Vu}\ and\ \citenamefont
  {Hasegawa}(2020)}]{Vu.2020.PRR}%
  \BibitemOpen
  \bibfield  {author} {\bibinfo {author} {\bibfnamefont {T.}~\bibnamefont
  {Van~Vu}}\ and\ \bibinfo {author} {\bibfnamefont {Y.}~\bibnamefont
  {Hasegawa}},\ }\bibfield  {title} {\bibinfo {title} {{Thermodynamic
  uncertainty relations under arbitrary control protocols}},\ }\href
  {https://doi.org/10.1103/PhysRevResearch.2.013060} {\bibfield  {journal}
  {\bibinfo  {journal} {Phys. Rev. Research}\ }\textbf {\bibinfo {volume}
  {2}},\ \bibinfo {pages} {013060} (\bibinfo {year} {2020})}\BibitemShut
  {NoStop}%
\bibitem [{\citenamefont {Koyuk}\ and\ \citenamefont
  {Seifert}(2020)}]{Koyuk.2020.PRL}%
  \BibitemOpen
  \bibfield  {author} {\bibinfo {author} {\bibfnamefont {T.}~\bibnamefont
  {Koyuk}}\ and\ \bibinfo {author} {\bibfnamefont {U.}~\bibnamefont
  {Seifert}},\ }\bibfield  {title} {\bibinfo {title} {{Thermodynamic
  uncertainty relation for time-dependent driving}},\ }\href
  {https://doi.org/10.1103/PhysRevLett.125.260604} {\bibfield  {journal}
  {\bibinfo  {journal} {Phys. Rev. Lett.}\ }\textbf {\bibinfo {volume} {125}},\
  \bibinfo {pages} {260604} (\bibinfo {year} {2020})}\BibitemShut {NoStop}%
\bibitem [{\citenamefont {Potts}\ and\ \citenamefont
  {Samuelsson}(2019)}]{Potts.2019.PRE}%
  \BibitemOpen
  \bibfield  {author} {\bibinfo {author} {\bibfnamefont {P.~P.}\ \bibnamefont
  {Potts}}\ and\ \bibinfo {author} {\bibfnamefont {P.}~\bibnamefont
  {Samuelsson}},\ }\bibfield  {title} {\bibinfo {title} {{Thermodynamic
  uncertainty relations including measurement and feedback}},\ }\href
  {https://doi.org/10.1103/PhysRevE.100.052137} {\bibfield  {journal} {\bibinfo
   {journal} {Phys. Rev. E}\ }\textbf {\bibinfo {volume} {100}},\ \bibinfo
  {pages} {052137} (\bibinfo {year} {2019})}\BibitemShut {NoStop}%
\bibitem [{\citenamefont {Vo}\ \emph {et~al.}(2020)\citenamefont {Vo},
  \citenamefont {Van~Vu},\ and\ \citenamefont {Hasegawa}}]{Vo.2020.PRE}%
  \BibitemOpen
  \bibfield  {author} {\bibinfo {author} {\bibfnamefont {V.~T.}\ \bibnamefont
  {Vo}}, \bibinfo {author} {\bibfnamefont {T.}~\bibnamefont {Van~Vu}},\ and\
  \bibinfo {author} {\bibfnamefont {Y.}~\bibnamefont {Hasegawa}},\ }\bibfield
  {title} {\bibinfo {title} {{Unified approach to classical speed limit and
  thermodynamic uncertainty relation}},\ }\href
  {https://doi.org/10.1103/PhysRevE.102.062132} {\bibfield  {journal} {\bibinfo
   {journal} {Phys. Rev. E}\ }\textbf {\bibinfo {volume} {102}},\ \bibinfo
  {pages} {062132} (\bibinfo {year} {2020})}\BibitemShut {NoStop}%
\bibitem [{\citenamefont {Falasco}\ \emph {et~al.}(2020)\citenamefont
  {Falasco}, \citenamefont {Esposito},\ and\ \citenamefont
  {Delvenne}}]{Falasco.2020.NJP}%
  \BibitemOpen
  \bibfield  {author} {\bibinfo {author} {\bibfnamefont {G.}~\bibnamefont
  {Falasco}}, \bibinfo {author} {\bibfnamefont {M.}~\bibnamefont {Esposito}},\
  and\ \bibinfo {author} {\bibfnamefont {J.-C.}\ \bibnamefont {Delvenne}},\
  }\bibfield  {title} {\bibinfo {title} {{Unifying thermodynamic uncertainty
  relations}},\ }\href {https://doi.org/10.1088/1367-2630/ab8679} {\bibfield
  {journal} {\bibinfo  {journal} {New J. Phys.}\ }\textbf {\bibinfo {volume}
  {22}},\ \bibinfo {pages} {053046} (\bibinfo {year} {2020})}\BibitemShut
  {NoStop}%
\bibitem [{\citenamefont {Friedman}\ \emph {et~al.}(2020)\citenamefont
  {Friedman}, \citenamefont {Agarwalla}, \citenamefont {Shein-Lumbroso},
  \citenamefont {Tal},\ and\ \citenamefont {Segal}}]{Friedman.2020.PRB}%
  \BibitemOpen
  \bibfield  {author} {\bibinfo {author} {\bibfnamefont {H.~M.}\ \bibnamefont
  {Friedman}}, \bibinfo {author} {\bibfnamefont {B.~K.}\ \bibnamefont
  {Agarwalla}}, \bibinfo {author} {\bibfnamefont {O.}~\bibnamefont
  {Shein-Lumbroso}}, \bibinfo {author} {\bibfnamefont {O.}~\bibnamefont
  {Tal}},\ and\ \bibinfo {author} {\bibfnamefont {D.}~\bibnamefont {Segal}},\
  }\bibfield  {title} {\bibinfo {title} {{Thermodynamic uncertainty relation in
  atomic-scale quantum conductors}},\ }\href
  {https://doi.org/10.1103/PhysRevB.101.195423} {\bibfield  {journal} {\bibinfo
   {journal} {Phys. Rev. B}\ }\textbf {\bibinfo {volume} {101}},\ \bibinfo
  {pages} {195423} (\bibinfo {year} {2020})}\BibitemShut {NoStop}%
\bibitem [{\citenamefont {Sacchi}(2021)}]{Sacchi.2021.PRE}%
  \BibitemOpen
  \bibfield  {author} {\bibinfo {author} {\bibfnamefont {M.~F.}\ \bibnamefont
  {Sacchi}},\ }\bibfield  {title} {\bibinfo {title} {{Thermodynamic uncertainty
  relations for bosonic Otto engines}},\ }\href
  {https://doi.org/10.1103/PhysRevE.103.012111} {\bibfield  {journal} {\bibinfo
   {journal} {Phys. Rev. E}\ }\textbf {\bibinfo {volume} {103}},\ \bibinfo
  {pages} {012111} (\bibinfo {year} {2021})}\BibitemShut {NoStop}%
\bibitem [{\citenamefont {Park}\ and\ \citenamefont
  {Park}(2021)}]{Park.2021.PRR}%
  \BibitemOpen
  \bibfield  {author} {\bibinfo {author} {\bibfnamefont {J.-M.}\ \bibnamefont
  {Park}}\ and\ \bibinfo {author} {\bibfnamefont {H.}~\bibnamefont {Park}},\
  }\bibfield  {title} {\bibinfo {title} {{Thermodynamic uncertainty relation in
  the overdamped limit with a magnetic Lorentz force}},\ }\href
  {https://doi.org/10.1103/PhysRevResearch.3.043005} {\bibfield  {journal}
  {\bibinfo  {journal} {Phys. Rev. Research}\ }\textbf {\bibinfo {volume}
  {3}},\ \bibinfo {pages} {043005} (\bibinfo {year} {2021})}\BibitemShut
  {NoStop}%
\bibitem [{\citenamefont {Lee}\ \emph {et~al.}(2021)\citenamefont {Lee},
  \citenamefont {Ha},\ and\ \citenamefont {Jeong}}]{Lee.2021.PRE}%
  \BibitemOpen
  \bibfield  {author} {\bibinfo {author} {\bibfnamefont {S.}~\bibnamefont
  {Lee}}, \bibinfo {author} {\bibfnamefont {M.}~\bibnamefont {Ha}},\ and\
  \bibinfo {author} {\bibfnamefont {H.}~\bibnamefont {Jeong}},\ }\bibfield
  {title} {\bibinfo {title} {{Quantumness and thermodynamic uncertainty
  relation of the finite-time Otto cycle}},\ }\href
  {https://doi.org/10.1103/PhysRevE.103.022136} {\bibfield  {journal} {\bibinfo
   {journal} {Phys. Rev. E}\ }\textbf {\bibinfo {volume} {103}},\ \bibinfo
  {pages} {022136} (\bibinfo {year} {2021})}\BibitemShut {NoStop}%
\bibitem [{\citenamefont {Dechant}\ and\ \citenamefont
  {Sasa}(2021)}]{Dechant.2021.PRR}%
  \BibitemOpen
  \bibfield  {author} {\bibinfo {author} {\bibfnamefont {A.}~\bibnamefont
  {Dechant}}\ and\ \bibinfo {author} {\bibfnamefont {S.-i.}\ \bibnamefont
  {Sasa}},\ }\bibfield  {title} {\bibinfo {title} {{Continuous time reversal
  and equality in the thermodynamic uncertainty relation}},\ }\href
  {https://doi.org/10.1103/PhysRevResearch.3.L042012} {\bibfield  {journal}
  {\bibinfo  {journal} {Phys. Rev. Research}\ }\textbf {\bibinfo {volume}
  {3}},\ \bibinfo {pages} {L042012} (\bibinfo {year} {2021})}\BibitemShut
  {NoStop}%
\bibitem [{\citenamefont {Timpanaro}\ \emph {et~al.}(2021)\citenamefont
  {Timpanaro}, \citenamefont {Guarnieri},\ and\ \citenamefont
  {Landi}}]{Timpanaro.2021.arxiv}%
  \BibitemOpen
  \bibfield  {author} {\bibinfo {author} {\bibfnamefont {A.~M.}\ \bibnamefont
  {Timpanaro}}, \bibinfo {author} {\bibfnamefont {G.}~\bibnamefont
  {Guarnieri}},\ and\ \bibinfo {author} {\bibfnamefont {G.~T.}\ \bibnamefont
  {Landi}},\ }\bibfield  {title} {\bibinfo {title} {{The most precise quantum
  thermoelectric}},\ }\href {https://arxiv.org/abs/2106.10205} {\bibfield
  {journal} {\bibinfo  {journal} {arXiv preprint arXiv:2106.10205}\ } (\bibinfo
  {year} {2021})}\BibitemShut {NoStop}%
\bibitem [{\citenamefont {Horodecki}\ and\ \citenamefont
  {Oppenheim}(2013)}]{Horodecki.2013.NC}%
  \BibitemOpen
  \bibfield  {author} {\bibinfo {author} {\bibfnamefont {M.}~\bibnamefont
  {Horodecki}}\ and\ \bibinfo {author} {\bibfnamefont {J.}~\bibnamefont
  {Oppenheim}},\ }\bibfield  {title} {\bibinfo {title} {{Fundamental
  limitations for quantum and nanoscale thermodynamics}},\ }\href
  {https://doi.org/10.1038/ncomms3059} {\bibfield  {journal} {\bibinfo
  {journal} {Nat. Commun.}\ }\textbf {\bibinfo {volume} {4}},\ \bibinfo {pages}
  {2059} (\bibinfo {year} {2013})}\BibitemShut {NoStop}%
\bibitem [{\citenamefont {Uzdin}\ \emph {et~al.}(2015)\citenamefont {Uzdin},
  \citenamefont {Levy},\ and\ \citenamefont {Kosloff}}]{Uzdin.2015.PRX}%
  \BibitemOpen
  \bibfield  {author} {\bibinfo {author} {\bibfnamefont {R.}~\bibnamefont
  {Uzdin}}, \bibinfo {author} {\bibfnamefont {A.}~\bibnamefont {Levy}},\ and\
  \bibinfo {author} {\bibfnamefont {R.}~\bibnamefont {Kosloff}},\ }\bibfield
  {title} {\bibinfo {title} {{Equivalence of quantum heat machines, and
  quantum-thermodynamic signatures}},\ }\href
  {https://doi.org/10.1103/PhysRevX.5.031044} {\bibfield  {journal} {\bibinfo
  {journal} {Phys. Rev. X}\ }\textbf {\bibinfo {volume} {5}},\ \bibinfo {pages}
  {031044} (\bibinfo {year} {2015})}\BibitemShut {NoStop}%
\bibitem [{\citenamefont {Lostaglio}\ \emph {et~al.}(2015)\citenamefont
  {Lostaglio}, \citenamefont {Korzekwa}, \citenamefont {Jennings},\ and\
  \citenamefont {Rudolph}}]{Lostaglio.2015.PRX}%
  \BibitemOpen
  \bibfield  {author} {\bibinfo {author} {\bibfnamefont {M.}~\bibnamefont
  {Lostaglio}}, \bibinfo {author} {\bibfnamefont {K.}~\bibnamefont {Korzekwa}},
  \bibinfo {author} {\bibfnamefont {D.}~\bibnamefont {Jennings}},\ and\
  \bibinfo {author} {\bibfnamefont {T.}~\bibnamefont {Rudolph}},\ }\bibfield
  {title} {\bibinfo {title} {{Quantum coherence, time-translation symmetry, and
  thermodynamics}},\ }\href {https://doi.org/10.1103/PhysRevX.5.021001}
  {\bibfield  {journal} {\bibinfo  {journal} {Phys. Rev. X}\ }\textbf {\bibinfo
  {volume} {5}},\ \bibinfo {pages} {021001} (\bibinfo {year}
  {2015})}\BibitemShut {NoStop}%
\bibitem [{\citenamefont {Korzekwa}\ \emph {et~al.}(2016)\citenamefont
  {Korzekwa}, \citenamefont {Lostaglio}, \citenamefont {Oppenheim},\ and\
  \citenamefont {Jennings}}]{Korzekwa.2016.NJP}%
  \BibitemOpen
  \bibfield  {author} {\bibinfo {author} {\bibfnamefont {K.}~\bibnamefont
  {Korzekwa}}, \bibinfo {author} {\bibfnamefont {M.}~\bibnamefont {Lostaglio}},
  \bibinfo {author} {\bibfnamefont {J.}~\bibnamefont {Oppenheim}},\ and\
  \bibinfo {author} {\bibfnamefont {D.}~\bibnamefont {Jennings}},\ }\bibfield
  {title} {\bibinfo {title} {{The extraction of work from quantum coherence}},\
  }\href {https://doi.org/10.1088/1367-2630/18/2/023045} {\bibfield  {journal}
  {\bibinfo  {journal} {New J. Phys.}\ }\textbf {\bibinfo {volume} {18}},\
  \bibinfo {pages} {023045} (\bibinfo {year} {2016})}\BibitemShut {NoStop}%
\bibitem [{\citenamefont {Francica}\ \emph {et~al.}(2019)\citenamefont
  {Francica}, \citenamefont {Goold},\ and\ \citenamefont
  {Plastina}}]{Francica.2019.PRE}%
  \BibitemOpen
  \bibfield  {author} {\bibinfo {author} {\bibfnamefont {G.}~\bibnamefont
  {Francica}}, \bibinfo {author} {\bibfnamefont {J.}~\bibnamefont {Goold}},\
  and\ \bibinfo {author} {\bibfnamefont {F.}~\bibnamefont {Plastina}},\
  }\bibfield  {title} {\bibinfo {title} {{Role of coherence in the
  nonequilibrium thermodynamics of quantum systems}},\ }\href
  {https://doi.org/10.1103/PhysRevE.99.042105} {\bibfield  {journal} {\bibinfo
  {journal} {Phys. Rev. E}\ }\textbf {\bibinfo {volume} {99}},\ \bibinfo
  {pages} {042105} (\bibinfo {year} {2019})}\BibitemShut {NoStop}%
\bibitem [{\citenamefont {Santos}\ \emph {et~al.}(2019)\citenamefont {Santos},
  \citenamefont {C{\'{e}}leri}, \citenamefont {Landi},\ and\ \citenamefont
  {Paternostro}}]{Santos.2019.npjQI}%
  \BibitemOpen
  \bibfield  {author} {\bibinfo {author} {\bibfnamefont {J.~P.}\ \bibnamefont
  {Santos}}, \bibinfo {author} {\bibfnamefont {L.~C.}\ \bibnamefont
  {C{\'{e}}leri}}, \bibinfo {author} {\bibfnamefont {G.~T.}\ \bibnamefont
  {Landi}},\ and\ \bibinfo {author} {\bibfnamefont {M.}~\bibnamefont
  {Paternostro}},\ }\bibfield  {title} {\bibinfo {title} {{The role of quantum
  coherence in non-equilibrium entropy production}},\ }\href
  {https://doi.org/10.1038/s41534-019-0138-y} {\bibfield  {journal} {\bibinfo
  {journal} {npj Quantum Inf.}\ }\textbf {\bibinfo {volume} {5}},\ \bibinfo
  {pages} {23} (\bibinfo {year} {2019})}\BibitemShut {NoStop}%
\bibitem [{\citenamefont {Francica}\ \emph {et~al.}(2020)\citenamefont
  {Francica}, \citenamefont {Binder}, \citenamefont {Guarnieri}, \citenamefont
  {Mitchison}, \citenamefont {Goold},\ and\ \citenamefont
  {Plastina}}]{Francica.2020.PRL}%
  \BibitemOpen
  \bibfield  {author} {\bibinfo {author} {\bibfnamefont {G.}~\bibnamefont
  {Francica}}, \bibinfo {author} {\bibfnamefont {F.~C.}\ \bibnamefont
  {Binder}}, \bibinfo {author} {\bibfnamefont {G.}~\bibnamefont {Guarnieri}},
  \bibinfo {author} {\bibfnamefont {M.~T.}\ \bibnamefont {Mitchison}}, \bibinfo
  {author} {\bibfnamefont {J.}~\bibnamefont {Goold}},\ and\ \bibinfo {author}
  {\bibfnamefont {F.}~\bibnamefont {Plastina}},\ }\bibfield  {title} {\bibinfo
  {title} {{Quantum coherence and ergotropy}},\ }\href
  {https://doi.org/10.1103/PhysRevLett.125.180603} {\bibfield  {journal}
  {\bibinfo  {journal} {Phys. Rev. Lett.}\ }\textbf {\bibinfo {volume} {125}},\
  \bibinfo {pages} {180603} (\bibinfo {year} {2020})}\BibitemShut {NoStop}%
\bibitem [{\citenamefont {Tajima}\ and\ \citenamefont
  {Funo}(2021)}]{Tajima.2021.PRL}%
  \BibitemOpen
  \bibfield  {author} {\bibinfo {author} {\bibfnamefont {H.}~\bibnamefont
  {Tajima}}\ and\ \bibinfo {author} {\bibfnamefont {K.}~\bibnamefont {Funo}},\
  }\bibfield  {title} {\bibinfo {title} {{Superconducting-like heat current:
  Effective cancellation of current-dissipation trade-off by quantum
  coherence}},\ }\href {https://doi.org/10.1103/PhysRevLett.127.190604}
  {\bibfield  {journal} {\bibinfo  {journal} {Phys. Rev. Lett.}\ }\textbf
  {\bibinfo {volume} {127}},\ \bibinfo {pages} {190604} (\bibinfo {year}
  {2021})}\BibitemShut {NoStop}%
\bibitem [{\citenamefont {Miller}\ \emph {et~al.}(2020)\citenamefont {Miller},
  \citenamefont {Guarnieri}, \citenamefont {Mitchison},\ and\ \citenamefont
  {Goold}}]{Miller.2020.PRL.QLP}%
  \BibitemOpen
  \bibfield  {author} {\bibinfo {author} {\bibfnamefont {H.~J.~D.}\
  \bibnamefont {Miller}}, \bibinfo {author} {\bibfnamefont {G.}~\bibnamefont
  {Guarnieri}}, \bibinfo {author} {\bibfnamefont {M.~T.}\ \bibnamefont
  {Mitchison}},\ and\ \bibinfo {author} {\bibfnamefont {J.}~\bibnamefont
  {Goold}},\ }\bibfield  {title} {\bibinfo {title} {{Quantum fluctuations
  hinder finite-time information erasure near the Landauer limit}},\ }\href
  {https://doi.org/10.1103/PhysRevLett.125.160602} {\bibfield  {journal}
  {\bibinfo  {journal} {Phys. Rev. Lett.}\ }\textbf {\bibinfo {volume} {125}},\
  \bibinfo {pages} {160602} (\bibinfo {year} {2020})}\BibitemShut {NoStop}%
\bibitem [{\citenamefont {Van~Vu}\ and\ \citenamefont
  {Saito}(2022)}]{Vu.2022.PRL}%
  \BibitemOpen
  \bibfield  {author} {\bibinfo {author} {\bibfnamefont {T.}~\bibnamefont
  {Van~Vu}}\ and\ \bibinfo {author} {\bibfnamefont {K.}~\bibnamefont {Saito}},\
  }\bibfield  {title} {\bibinfo {title} {{Finite-time quantum Landauer
  principle and quantum coherence}},\ }\href
  {https://doi.org/10.1103/PhysRevLett.128.010602} {\bibfield  {journal}
  {\bibinfo  {journal} {Phys. Rev. Lett.}\ }\textbf {\bibinfo {volume} {128}},\
  \bibinfo {pages} {010602} (\bibinfo {year} {2022})}\BibitemShut {NoStop}%
\bibitem [{\citenamefont {Brandner}\ \emph {et~al.}(2017)\citenamefont
  {Brandner}, \citenamefont {Bauer},\ and\ \citenamefont
  {Seifert}}]{Brandner.2017.PRL}%
  \BibitemOpen
  \bibfield  {author} {\bibinfo {author} {\bibfnamefont {K.}~\bibnamefont
  {Brandner}}, \bibinfo {author} {\bibfnamefont {M.}~\bibnamefont {Bauer}},\
  and\ \bibinfo {author} {\bibfnamefont {U.}~\bibnamefont {Seifert}},\
  }\bibfield  {title} {\bibinfo {title} {{Universal coherence-induced power
  losses of quantum heat engines in linear response}},\ }\href
  {https://doi.org/10.1103/PhysRevLett.119.170602} {\bibfield  {journal}
  {\bibinfo  {journal} {Phys. Rev. Lett.}\ }\textbf {\bibinfo {volume} {119}},\
  \bibinfo {pages} {170602} (\bibinfo {year} {2017})}\BibitemShut {NoStop}%
\bibitem [{\citenamefont {Brandner}\ and\ \citenamefont
  {Saito}(2020)}]{Brandner.2020.PRL}%
  \BibitemOpen
  \bibfield  {author} {\bibinfo {author} {\bibfnamefont {K.}~\bibnamefont
  {Brandner}}\ and\ \bibinfo {author} {\bibfnamefont {K.}~\bibnamefont
  {Saito}},\ }\bibfield  {title} {\bibinfo {title} {{Thermodynamic geometry of
  microscopic heat engines}},\ }\href
  {https://doi.org/10.1103/PhysRevLett.124.040602} {\bibfield  {journal}
  {\bibinfo  {journal} {Phys. Rev. Lett.}\ }\textbf {\bibinfo {volume} {124}},\
  \bibinfo {pages} {040602} (\bibinfo {year} {2020})}\BibitemShut {NoStop}%
\bibitem [{\citenamefont {Scully}\ \emph {et~al.}(2011)\citenamefont {Scully},
  \citenamefont {Chapin}, \citenamefont {Dorfman}, \citenamefont {Kim},\ and\
  \citenamefont {Svidzinsky}}]{Scully.2011.PNAS}%
  \BibitemOpen
  \bibfield  {author} {\bibinfo {author} {\bibfnamefont {M.~O.}\ \bibnamefont
  {Scully}}, \bibinfo {author} {\bibfnamefont {K.~R.}\ \bibnamefont {Chapin}},
  \bibinfo {author} {\bibfnamefont {K.~E.}\ \bibnamefont {Dorfman}}, \bibinfo
  {author} {\bibfnamefont {M.~B.}\ \bibnamefont {Kim}},\ and\ \bibinfo {author}
  {\bibfnamefont {A.}~\bibnamefont {Svidzinsky}},\ }\bibfield  {title}
  {\bibinfo {title} {{Quantum heat engine power can be increased by
  noise-induced coherence}},\ }\href {https://doi.org/10.1073/pnas.1110234108}
  {\bibfield  {journal} {\bibinfo  {journal} {Proc. Natl. Acad. Sci. U.S.A.}\
  }\textbf {\bibinfo {volume} {108}},\ \bibinfo {pages} {15097} (\bibinfo
  {year} {2011})}\BibitemShut {NoStop}%
\bibitem [{\citenamefont {Watanabe}\ \emph {et~al.}(2017)\citenamefont
  {Watanabe}, \citenamefont {Venkatesh}, \citenamefont {Talkner},\ and\
  \citenamefont {del Campo}}]{Watanabe.2017.PRL}%
  \BibitemOpen
  \bibfield  {author} {\bibinfo {author} {\bibfnamefont {G.}~\bibnamefont
  {Watanabe}}, \bibinfo {author} {\bibfnamefont {B.~P.}\ \bibnamefont
  {Venkatesh}}, \bibinfo {author} {\bibfnamefont {P.}~\bibnamefont {Talkner}},\
  and\ \bibinfo {author} {\bibfnamefont {A.}~\bibnamefont {del Campo}},\
  }\bibfield  {title} {\bibinfo {title} {{Quantum performance of thermal
  machines over many cycles}},\ }\href
  {https://doi.org/10.1103/PhysRevLett.118.050601} {\bibfield  {journal}
  {\bibinfo  {journal} {Phys. Rev. Lett.}\ }\textbf {\bibinfo {volume} {118}},\
  \bibinfo {pages} {050601} (\bibinfo {year} {2017})}\BibitemShut {NoStop}%
\bibitem [{\citenamefont {Menczel}\ \emph {et~al.}(2020)\citenamefont
  {Menczel}, \citenamefont {Flindt},\ and\ \citenamefont
  {Brandner}}]{Menczel.2020.PRA}%
  \BibitemOpen
  \bibfield  {author} {\bibinfo {author} {\bibfnamefont {P.}~\bibnamefont
  {Menczel}}, \bibinfo {author} {\bibfnamefont {C.}~\bibnamefont {Flindt}},\
  and\ \bibinfo {author} {\bibfnamefont {K.}~\bibnamefont {Brandner}},\
  }\bibfield  {title} {\bibinfo {title} {{Thermodynamics of cyclic quantum
  amplifiers}},\ }\href {https://doi.org/10.1103/PhysRevA.101.052106}
  {\bibfield  {journal} {\bibinfo  {journal} {Phys. Rev. A}\ }\textbf {\bibinfo
  {volume} {101}},\ \bibinfo {pages} {052106} (\bibinfo {year}
  {2020})}\BibitemShut {NoStop}%
\bibitem [{\citenamefont {Ptaszy\ifmmode~\acute{n}\else
  \'{n}\fi{}ski}(2018)}]{Ptaszynski.2018.PRB}%
  \BibitemOpen
  \bibfield  {author} {\bibinfo {author} {\bibfnamefont {K.}~\bibnamefont
  {Ptaszy\ifmmode~\acute{n}\else \'{n}\fi{}ski}},\ }\bibfield  {title}
  {\bibinfo {title} {{Coherence-enhanced constancy of a quantum thermoelectric
  generator}},\ }\href {https://doi.org/10.1103/PhysRevB.98.085425} {\bibfield
  {journal} {\bibinfo  {journal} {Phys. Rev. B}\ }\textbf {\bibinfo {volume}
  {98}},\ \bibinfo {pages} {085425} (\bibinfo {year} {2018})}\BibitemShut
  {NoStop}%
\bibitem [{\citenamefont {Cangemi}\ \emph {et~al.}(2020)\citenamefont
  {Cangemi}, \citenamefont {Cataudella}, \citenamefont {Benenti}, \citenamefont
  {Sassetti},\ and\ \citenamefont {De~Filippis}}]{Cangemi.2020.PRB}%
  \BibitemOpen
  \bibfield  {author} {\bibinfo {author} {\bibfnamefont {L.~M.}\ \bibnamefont
  {Cangemi}}, \bibinfo {author} {\bibfnamefont {V.}~\bibnamefont {Cataudella}},
  \bibinfo {author} {\bibfnamefont {G.}~\bibnamefont {Benenti}}, \bibinfo
  {author} {\bibfnamefont {M.}~\bibnamefont {Sassetti}},\ and\ \bibinfo
  {author} {\bibfnamefont {G.}~\bibnamefont {De~Filippis}},\ }\bibfield
  {title} {\bibinfo {title} {{Violation of thermodynamics uncertainty relations
  in a periodically driven work-to-work converter from weak to strong
  dissipation}},\ }\href {https://doi.org/10.1103/PhysRevB.102.165418}
  {\bibfield  {journal} {\bibinfo  {journal} {Phys. Rev. B}\ }\textbf {\bibinfo
  {volume} {102}},\ \bibinfo {pages} {165418} (\bibinfo {year}
  {2020})}\BibitemShut {NoStop}%
\bibitem [{\citenamefont {Kalaee}\ \emph {et~al.}(2021)\citenamefont {Kalaee},
  \citenamefont {Wacker},\ and\ \citenamefont {Potts}}]{Kalaee.2021.PRE}%
  \BibitemOpen
  \bibfield  {author} {\bibinfo {author} {\bibfnamefont {A.~A.~S.}\
  \bibnamefont {Kalaee}}, \bibinfo {author} {\bibfnamefont {A.}~\bibnamefont
  {Wacker}},\ and\ \bibinfo {author} {\bibfnamefont {P.~P.}\ \bibnamefont
  {Potts}},\ }\bibfield  {title} {\bibinfo {title} {{Violating the
  thermodynamic uncertainty relation in the three-level maser}},\ }\href
  {https://doi.org/10.1103/PhysRevE.104.L012103} {\bibfield  {journal}
  {\bibinfo  {journal} {Phys. Rev. E}\ }\textbf {\bibinfo {volume} {104}},\
  \bibinfo {pages} {L012103} (\bibinfo {year} {2021})}\BibitemShut {NoStop}%
\bibitem [{\citenamefont {Menczel}\ \emph {et~al.}(2021)\citenamefont
  {Menczel}, \citenamefont {Loisa}, \citenamefont {Brandner},\ and\
  \citenamefont {Flindt}}]{Menczel.2021.JPA}%
  \BibitemOpen
  \bibfield  {author} {\bibinfo {author} {\bibfnamefont {P.}~\bibnamefont
  {Menczel}}, \bibinfo {author} {\bibfnamefont {E.}~\bibnamefont {Loisa}},
  \bibinfo {author} {\bibfnamefont {K.}~\bibnamefont {Brandner}},\ and\
  \bibinfo {author} {\bibfnamefont {C.}~\bibnamefont {Flindt}},\ }\bibfield
  {title} {\bibinfo {title} {{Thermodynamic uncertainty relations for
  coherently driven open quantum systems}},\ }\href
  {https://doi.org/10.1088/1751-8121/ac0c8f} {\bibfield  {journal} {\bibinfo
  {journal} {J. Phys. A}\ }\textbf {\bibinfo {volume} {54}},\ \bibinfo {pages}
  {314002} (\bibinfo {year} {2021})}\BibitemShut {NoStop}%
\bibitem [{\citenamefont {Rignon-Bret}\ \emph {et~al.}(2021)\citenamefont
  {Rignon-Bret}, \citenamefont {Guarnieri}, \citenamefont {Goold},\ and\
  \citenamefont {Mitchison}}]{Bret.2021.PRE}%
  \BibitemOpen
  \bibfield  {author} {\bibinfo {author} {\bibfnamefont {A.}~\bibnamefont
  {Rignon-Bret}}, \bibinfo {author} {\bibfnamefont {G.}~\bibnamefont
  {Guarnieri}}, \bibinfo {author} {\bibfnamefont {J.}~\bibnamefont {Goold}},\
  and\ \bibinfo {author} {\bibfnamefont {M.~T.}\ \bibnamefont {Mitchison}},\
  }\bibfield  {title} {\bibinfo {title} {{Thermodynamics of precision in
  quantum nanomachines}},\ }\href {https://doi.org/10.1103/PhysRevE.103.012133}
  {\bibfield  {journal} {\bibinfo  {journal} {Phys. Rev. E}\ }\textbf {\bibinfo
  {volume} {103}},\ \bibinfo {pages} {012133} (\bibinfo {year}
  {2021})}\BibitemShut {NoStop}%
\bibitem [{\citenamefont {Erker}\ \emph {et~al.}(2017)\citenamefont {Erker},
  \citenamefont {Mitchison}, \citenamefont {Silva}, \citenamefont {Woods},
  \citenamefont {Brunner},\ and\ \citenamefont {Huber}}]{Erker.2017.PRX}%
  \BibitemOpen
  \bibfield  {author} {\bibinfo {author} {\bibfnamefont {P.}~\bibnamefont
  {Erker}}, \bibinfo {author} {\bibfnamefont {M.~T.}\ \bibnamefont
  {Mitchison}}, \bibinfo {author} {\bibfnamefont {R.}~\bibnamefont {Silva}},
  \bibinfo {author} {\bibfnamefont {M.~P.}\ \bibnamefont {Woods}}, \bibinfo
  {author} {\bibfnamefont {N.}~\bibnamefont {Brunner}},\ and\ \bibinfo {author}
  {\bibfnamefont {M.}~\bibnamefont {Huber}},\ }\bibfield  {title} {\bibinfo
  {title} {{Autonomous quantum clocks: Does thermodynamics limit our ability to
  measure time?}},\ }\href {https://doi.org/10.1103/PhysRevX.7.031022}
  {\bibfield  {journal} {\bibinfo  {journal} {Phys. Rev. X}\ }\textbf {\bibinfo
  {volume} {7}},\ \bibinfo {pages} {031022} (\bibinfo {year}
  {2017})}\BibitemShut {NoStop}%
\bibitem [{\citenamefont {Guarnieri}\ \emph {et~al.}(2019)\citenamefont
  {Guarnieri}, \citenamefont {Landi}, \citenamefont {Clark},\ and\
  \citenamefont {Goold}}]{Guarnieri.2019.PRR}%
  \BibitemOpen
  \bibfield  {author} {\bibinfo {author} {\bibfnamefont {G.}~\bibnamefont
  {Guarnieri}}, \bibinfo {author} {\bibfnamefont {G.~T.}\ \bibnamefont
  {Landi}}, \bibinfo {author} {\bibfnamefont {S.~R.}\ \bibnamefont {Clark}},\
  and\ \bibinfo {author} {\bibfnamefont {J.}~\bibnamefont {Goold}},\ }\bibfield
   {title} {\bibinfo {title} {{Thermodynamics of precision in quantum
  nonequilibrium steady states}},\ }\href
  {https://doi.org/10.1103/PhysRevResearch.1.033021} {\bibfield  {journal}
  {\bibinfo  {journal} {Phys. Rev. Research}\ }\textbf {\bibinfo {volume}
  {1}},\ \bibinfo {pages} {033021} (\bibinfo {year} {2019})}\BibitemShut
  {NoStop}%
\bibitem [{\citenamefont {Carollo}\ \emph {et~al.}(2019)\citenamefont
  {Carollo}, \citenamefont {Jack},\ and\ \citenamefont
  {Garrahan}}]{Carollo.2019.PRL}%
  \BibitemOpen
  \bibfield  {author} {\bibinfo {author} {\bibfnamefont {F.}~\bibnamefont
  {Carollo}}, \bibinfo {author} {\bibfnamefont {R.~L.}\ \bibnamefont {Jack}},\
  and\ \bibinfo {author} {\bibfnamefont {J.~P.}\ \bibnamefont {Garrahan}},\
  }\bibfield  {title} {\bibinfo {title} {{Unraveling the large deviation
  statistics of Markovian open quantum systems}},\ }\href
  {https://doi.org/10.1103/PhysRevLett.122.130605} {\bibfield  {journal}
  {\bibinfo  {journal} {Phys. Rev. Lett.}\ }\textbf {\bibinfo {volume} {122}},\
  \bibinfo {pages} {130605} (\bibinfo {year} {2019})}\BibitemShut {NoStop}%
\bibitem [{\citenamefont {Hasegawa}(2020)}]{Hasegawa.2020.PRL}%
  \BibitemOpen
  \bibfield  {author} {\bibinfo {author} {\bibfnamefont {Y.}~\bibnamefont
  {Hasegawa}},\ }\bibfield  {title} {\bibinfo {title} {{Quantum thermodynamic
  uncertainty relation for continuous measurement}},\ }\href
  {https://doi.org/10.1103/PhysRevLett.125.050601} {\bibfield  {journal}
  {\bibinfo  {journal} {Phys. Rev. Lett.}\ }\textbf {\bibinfo {volume} {125}},\
  \bibinfo {pages} {050601} (\bibinfo {year} {2020})}\BibitemShut {NoStop}%
\bibitem [{\citenamefont {Hasegawa}(2021{\natexlab{a}})}]{Hasegawa.2021.PRL}%
  \BibitemOpen
  \bibfield  {author} {\bibinfo {author} {\bibfnamefont {Y.}~\bibnamefont
  {Hasegawa}},\ }\bibfield  {title} {\bibinfo {title} {{Thermodynamic
  uncertainty relation for general open quantum systems}},\ }\href
  {https://doi.org/10.1103/PhysRevLett.126.010602} {\bibfield  {journal}
  {\bibinfo  {journal} {Phys. Rev. Lett.}\ }\textbf {\bibinfo {volume} {126}},\
  \bibinfo {pages} {010602} (\bibinfo {year} {2021}{\natexlab{a}})}\BibitemShut
  {NoStop}%
\bibitem [{\citenamefont {Miller}\ \emph
  {et~al.}(2021{\natexlab{a}})\citenamefont {Miller}, \citenamefont
  {Mohammady}, \citenamefont {Perarnau-Llobet},\ and\ \citenamefont
  {Guarnieri}}]{Miller.2021.PRL.TUR}%
  \BibitemOpen
  \bibfield  {author} {\bibinfo {author} {\bibfnamefont {H.~J.~D.}\
  \bibnamefont {Miller}}, \bibinfo {author} {\bibfnamefont {M.~H.}\
  \bibnamefont {Mohammady}}, \bibinfo {author} {\bibfnamefont {M.}~\bibnamefont
  {Perarnau-Llobet}},\ and\ \bibinfo {author} {\bibfnamefont {G.}~\bibnamefont
  {Guarnieri}},\ }\bibfield  {title} {\bibinfo {title} {{Thermodynamic
  uncertainty relation in slowly driven quantum heat engines}},\ }\href
  {https://doi.org/10.1103/PhysRevLett.126.210603} {\bibfield  {journal}
  {\bibinfo  {journal} {Phys. Rev. Lett.}\ }\textbf {\bibinfo {volume} {126}},\
  \bibinfo {pages} {210603} (\bibinfo {year} {2021}{\natexlab{a}})}\BibitemShut
  {NoStop}%
\bibitem [{\citenamefont {Hasegawa}(2021{\natexlab{b}})}]{Hasegawa.2021.PRL2}%
  \BibitemOpen
  \bibfield  {author} {\bibinfo {author} {\bibfnamefont {Y.}~\bibnamefont
  {Hasegawa}},\ }\bibfield  {title} {\bibinfo {title} {{Irreversibility,
  Loschmidt echo, and thermodynamic uncertainty relation}},\ }\href
  {https://doi.org/10.1103/PhysRevLett.127.240602} {\bibfield  {journal}
  {\bibinfo  {journal} {Phys. Rev. Lett.}\ }\textbf {\bibinfo {volume} {127}},\
  \bibinfo {pages} {240602} (\bibinfo {year} {2021}{\natexlab{b}})}\BibitemShut
  {NoStop}%
\bibitem [{\citenamefont {Streltsov}\ \emph {et~al.}(2017)\citenamefont
  {Streltsov}, \citenamefont {Adesso},\ and\ \citenamefont
  {Plenio}}]{Streltsov.2017.RMP}%
  \BibitemOpen
  \bibfield  {author} {\bibinfo {author} {\bibfnamefont {A.}~\bibnamefont
  {Streltsov}}, \bibinfo {author} {\bibfnamefont {G.}~\bibnamefont {Adesso}},\
  and\ \bibinfo {author} {\bibfnamefont {M.~B.}\ \bibnamefont {Plenio}},\
  }\bibfield  {title} {\bibinfo {title} {{Colloquium: Quantum coherence as a
  resource}},\ }\href {https://doi.org/10.1103/RevModPhys.89.041003} {\bibfield
   {journal} {\bibinfo  {journal} {Rev. Mod. Phys.}\ }\textbf {\bibinfo
  {volume} {89}},\ \bibinfo {pages} {041003} (\bibinfo {year}
  {2017})}\BibitemShut {NoStop}%
\bibitem [{\citenamefont {Scovil}\ and\ \citenamefont
  {Schulz-DuBois}(1959)}]{Scovil.1959.PRL}%
  \BibitemOpen
  \bibfield  {author} {\bibinfo {author} {\bibfnamefont {H.~E.~D.}\
  \bibnamefont {Scovil}}\ and\ \bibinfo {author} {\bibfnamefont {E.~O.}\
  \bibnamefont {Schulz-DuBois}},\ }\bibfield  {title} {\bibinfo {title}
  {{Three-level masers as heat engines}},\ }\href
  {https://doi.org/10.1103/PhysRevLett.2.262} {\bibfield  {journal} {\bibinfo
  {journal} {Phys. Rev. Lett.}\ }\textbf {\bibinfo {volume} {2}},\ \bibinfo
  {pages} {262} (\bibinfo {year} {1959})}\BibitemShut {NoStop}%
\bibitem [{\citenamefont {Zou}\ \emph {et~al.}(2017)\citenamefont {Zou},
  \citenamefont {Jiang}, \citenamefont {Mei}, \citenamefont {Guo},\ and\
  \citenamefont {Du}}]{Zou.2017.PRL}%
  \BibitemOpen
  \bibfield  {author} {\bibinfo {author} {\bibfnamefont {Y.}~\bibnamefont
  {Zou}}, \bibinfo {author} {\bibfnamefont {Y.}~\bibnamefont {Jiang}}, \bibinfo
  {author} {\bibfnamefont {Y.}~\bibnamefont {Mei}}, \bibinfo {author}
  {\bibfnamefont {X.}~\bibnamefont {Guo}},\ and\ \bibinfo {author}
  {\bibfnamefont {S.}~\bibnamefont {Du}},\ }\bibfield  {title} {\bibinfo
  {title} {{Quantum heat engine using electromagnetically induced
  transparency}},\ }\href {https://doi.org/10.1103/PhysRevLett.119.050602}
  {\bibfield  {journal} {\bibinfo  {journal} {Phys. Rev. Lett.}\ }\textbf
  {\bibinfo {volume} {119}},\ \bibinfo {pages} {050602} (\bibinfo {year}
  {2017})}\BibitemShut {NoStop}%
\bibitem [{\citenamefont {Klatzow}\ \emph {et~al.}(2019)\citenamefont
  {Klatzow}, \citenamefont {Becker}, \citenamefont {Ledingham}, \citenamefont
  {Weinzetl}, \citenamefont {Kaczmarek}, \citenamefont {Saunders},
  \citenamefont {Nunn}, \citenamefont {Walmsley}, \citenamefont {Uzdin},\ and\
  \citenamefont {Poem}}]{Klatzow.2019.PRL}%
  \BibitemOpen
  \bibfield  {author} {\bibinfo {author} {\bibfnamefont {J.}~\bibnamefont
  {Klatzow}}, \bibinfo {author} {\bibfnamefont {J.~N.}\ \bibnamefont {Becker}},
  \bibinfo {author} {\bibfnamefont {P.~M.}\ \bibnamefont {Ledingham}}, \bibinfo
  {author} {\bibfnamefont {C.}~\bibnamefont {Weinzetl}}, \bibinfo {author}
  {\bibfnamefont {K.~T.}\ \bibnamefont {Kaczmarek}}, \bibinfo {author}
  {\bibfnamefont {D.~J.}\ \bibnamefont {Saunders}}, \bibinfo {author}
  {\bibfnamefont {J.}~\bibnamefont {Nunn}}, \bibinfo {author} {\bibfnamefont
  {I.~A.}\ \bibnamefont {Walmsley}}, \bibinfo {author} {\bibfnamefont
  {R.}~\bibnamefont {Uzdin}},\ and\ \bibinfo {author} {\bibfnamefont
  {E.}~\bibnamefont {Poem}},\ }\bibfield  {title} {\bibinfo {title}
  {{Experimental demonstration of quantum effects in the operation of
  microscopic heat engines}},\ }\href
  {https://doi.org/10.1103/PhysRevLett.122.110601} {\bibfield  {journal}
  {\bibinfo  {journal} {Phys. Rev. Lett.}\ }\textbf {\bibinfo {volume} {122}},\
  \bibinfo {pages} {110601} (\bibinfo {year} {2019})}\BibitemShut {NoStop}%
\bibitem [{\citenamefont {Lindblad}(1976)}]{Lindblad.1976.CMP}%
  \BibitemOpen
  \bibfield  {author} {\bibinfo {author} {\bibfnamefont {G.}~\bibnamefont
  {Lindblad}},\ }\bibfield  {title} {\bibinfo {title} {{On the generators of
  quantum dynamical semigroups}},\ }\href {https://doi.org/10.1007/BF01608499}
  {\bibfield  {journal} {\bibinfo  {journal} {Commun. Math. Phys.}\ }\textbf
  {\bibinfo {volume} {48}},\ \bibinfo {pages} {119} (\bibinfo {year}
  {1976})}\BibitemShut {NoStop}%
\bibitem [{\citenamefont {Gorini}\ \emph {et~al.}(1976)\citenamefont {Gorini},
  \citenamefont {Kossakowski},\ and\ \citenamefont
  {Sudarshan}}]{Gorini.1976.JMP}%
  \BibitemOpen
  \bibfield  {author} {\bibinfo {author} {\bibfnamefont {V.}~\bibnamefont
  {Gorini}}, \bibinfo {author} {\bibfnamefont {A.}~\bibnamefont
  {Kossakowski}},\ and\ \bibinfo {author} {\bibfnamefont {E.~C.~G.}\
  \bibnamefont {Sudarshan}},\ }\bibfield  {title} {\bibinfo {title}
  {{Completely positive dynamical semigroups of N‐level systems}},\ }\href
  {https://doi.org/10.1063/1.522979} {\bibfield  {journal} {\bibinfo  {journal}
  {J. Math. Phys.}\ }\textbf {\bibinfo {volume} {17}},\ \bibinfo {pages} {821}
  (\bibinfo {year} {1976})}\BibitemShut {NoStop}%
\bibitem [{\citenamefont {Wiseman}\ and\ \citenamefont
  {Milburn}(2009)}]{Wiseman.2009}%
  \BibitemOpen
  \bibfield  {author} {\bibinfo {author} {\bibfnamefont {H.~M.}\ \bibnamefont
  {Wiseman}}\ and\ \bibinfo {author} {\bibfnamefont {G.~J.}\ \bibnamefont
  {Milburn}},\ }\href@noop {} {\emph {\bibinfo {title} {{Quantum Measurement
  and Control}}}}\ (\bibinfo  {publisher} {Cambridge University Press},\
  \bibinfo {address} {Cambridge},\ \bibinfo {year} {2009})\BibitemShut
  {NoStop}%
\bibitem [{\citenamefont {Manzano}\ \emph {et~al.}(2019)\citenamefont
  {Manzano}, \citenamefont {Fazio},\ and\ \citenamefont
  {Rold\'an}}]{Manzano.2019.PRL}%
  \BibitemOpen
  \bibfield  {author} {\bibinfo {author} {\bibfnamefont {G.}~\bibnamefont
  {Manzano}}, \bibinfo {author} {\bibfnamefont {R.}~\bibnamefont {Fazio}},\
  and\ \bibinfo {author} {\bibfnamefont {E.}~\bibnamefont {Rold\'an}},\
  }\bibfield  {title} {\bibinfo {title} {{Quantum martingale theory and entropy
  production}},\ }\href {https://doi.org/10.1103/PhysRevLett.122.220602}
  {\bibfield  {journal} {\bibinfo  {journal} {Phys. Rev. Lett.}\ }\textbf
  {\bibinfo {volume} {122}},\ \bibinfo {pages} {220602} (\bibinfo {year}
  {2019})}\BibitemShut {NoStop}%
\bibitem [{\citenamefont {Horowitz}(2012)}]{Horowitz.2012.PRE}%
  \BibitemOpen
  \bibfield  {author} {\bibinfo {author} {\bibfnamefont {J.~M.}\ \bibnamefont
  {Horowitz}},\ }\bibfield  {title} {\bibinfo {title} {{Quantum-trajectory
  approach to the stochastic thermodynamics of a forced harmonic oscillator}},\
  }\href {https://doi.org/10.1103/PhysRevE.85.031110} {\bibfield  {journal}
  {\bibinfo  {journal} {Phys. Rev. E}\ }\textbf {\bibinfo {volume} {85}},\
  \bibinfo {pages} {031110} (\bibinfo {year} {2012})}\BibitemShut {NoStop}%
\bibitem [{\citenamefont {Horowitz}\ and\ \citenamefont
  {Parrondo}(2013)}]{Horowitz.2013.NJP}%
  \BibitemOpen
  \bibfield  {author} {\bibinfo {author} {\bibfnamefont {J.~M.}\ \bibnamefont
  {Horowitz}}\ and\ \bibinfo {author} {\bibfnamefont {J.~M.~R.}\ \bibnamefont
  {Parrondo}},\ }\bibfield  {title} {\bibinfo {title} {{Entropy production
  along nonequilibrium quantum jump trajectories}},\ }\href
  {https://doi.org/10.1088/1367-2630/15/8/085028} {\bibfield  {journal}
  {\bibinfo  {journal} {New J. Phys.}\ }\textbf {\bibinfo {volume} {15}},\
  \bibinfo {pages} {085028} (\bibinfo {year} {2013})}\BibitemShut {NoStop}%
\bibitem [{\citenamefont {Manzano}\ \emph {et~al.}(2015)\citenamefont
  {Manzano}, \citenamefont {Horowitz},\ and\ \citenamefont
  {Parrondo}}]{Manzano.2015.PRE}%
  \BibitemOpen
  \bibfield  {author} {\bibinfo {author} {\bibfnamefont {G.}~\bibnamefont
  {Manzano}}, \bibinfo {author} {\bibfnamefont {J.~M.}\ \bibnamefont
  {Horowitz}},\ and\ \bibinfo {author} {\bibfnamefont {J.~M.~R.}\ \bibnamefont
  {Parrondo}},\ }\bibfield  {title} {\bibinfo {title} {{Nonequilibrium
  potential and fluctuation theorems for quantum maps}},\ }\href
  {https://doi.org/10.1103/PhysRevE.92.032129} {\bibfield  {journal} {\bibinfo
  {journal} {Phys. Rev. E}\ }\textbf {\bibinfo {volume} {92}},\ \bibinfo
  {pages} {032129} (\bibinfo {year} {2015})}\BibitemShut {NoStop}%
\bibitem [{\citenamefont {Miller}\ \emph
  {et~al.}(2021{\natexlab{b}})\citenamefont {Miller}, \citenamefont
  {Mohammady}, \citenamefont {Perarnau-Llobet},\ and\ \citenamefont
  {Guarnieri}}]{Miller.2021.PRE}%
  \BibitemOpen
  \bibfield  {author} {\bibinfo {author} {\bibfnamefont {H.~J.~D.}\
  \bibnamefont {Miller}}, \bibinfo {author} {\bibfnamefont {M.~H.}\
  \bibnamefont {Mohammady}}, \bibinfo {author} {\bibfnamefont {M.}~\bibnamefont
  {Perarnau-Llobet}},\ and\ \bibinfo {author} {\bibfnamefont {G.}~\bibnamefont
  {Guarnieri}},\ }\bibfield  {title} {\bibinfo {title} {{Joint statistics of
  work and entropy production along quantum trajectories}},\ }\href
  {https://doi.org/10.1103/PhysRevE.103.052138} {\bibfield  {journal} {\bibinfo
   {journal} {Phys. Rev. E}\ }\textbf {\bibinfo {volume} {103}},\ \bibinfo
  {pages} {052138} (\bibinfo {year} {2021}{\natexlab{b}})}\BibitemShut
  {NoStop}%
\bibitem [{\citenamefont {Breuer}\ and\ \citenamefont
  {Petruccione}(2002)}]{Breuer.2002}%
  \BibitemOpen
  \bibfield  {author} {\bibinfo {author} {\bibfnamefont {H.-P.}\ \bibnamefont
  {Breuer}}\ and\ \bibinfo {author} {\bibfnamefont {F.}~\bibnamefont
  {Petruccione}},\ }\href@noop {} {\emph {\bibinfo {title} {{The Theory of Open
  Quantum Systems}}}}\ (\bibinfo  {publisher} {Oxford University Press},\
  \bibinfo {address} {New York},\ \bibinfo {year} {2002})\BibitemShut {NoStop}%
\bibitem [{fnt()}]{fnt0}%
  \BibitemOpen
  \href@noop {} {}For the sake of unified bounds for both observables $\Phi$
  and $\Lambda$, we have employed the rescaling by time in
  Eq.~\eqref{eq:static.obs}. If we do not rescale, the bound
  \eqref{eq:main.result.2} will be slightly different from the current
  form.\BibitemShut {Stop}%
\bibitem [{\citenamefont {Landi}\ and\ \citenamefont
  {Paternostro}(2021)}]{Landi.2021.RMP}%
  \BibitemOpen
  \bibfield  {author} {\bibinfo {author} {\bibfnamefont {G.~T.}\ \bibnamefont
  {Landi}}\ and\ \bibinfo {author} {\bibfnamefont {M.}~\bibnamefont
  {Paternostro}},\ }\bibfield  {title} {\bibinfo {title} {{Irreversible entropy
  production: From classical to quantum}},\ }\href
  {https://doi.org/10.1103/RevModPhys.93.035008} {\bibfield  {journal}
  {\bibinfo  {journal} {Rev. Mod. Phys.}\ }\textbf {\bibinfo {volume} {93}},\
  \bibinfo {pages} {035008} (\bibinfo {year} {2021})}\BibitemShut {NoStop}%
\bibitem [{\citenamefont {Shiraishi}\ \emph {et~al.}(2018)\citenamefont
  {Shiraishi}, \citenamefont {Funo},\ and\ \citenamefont
  {Saito}}]{Shiraishi.2018.PRL}%
  \BibitemOpen
  \bibfield  {author} {\bibinfo {author} {\bibfnamefont {N.}~\bibnamefont
  {Shiraishi}}, \bibinfo {author} {\bibfnamefont {K.}~\bibnamefont {Funo}},\
  and\ \bibinfo {author} {\bibfnamefont {K.}~\bibnamefont {Saito}},\ }\bibfield
   {title} {\bibinfo {title} {{Speed limit for classical stochastic
  processes}},\ }\href {https://doi.org/10.1103/PhysRevLett.121.070601}
  {\bibfield  {journal} {\bibinfo  {journal} {Phys. Rev. Lett.}\ }\textbf
  {\bibinfo {volume} {121}},\ \bibinfo {pages} {070601} (\bibinfo {year}
  {2018})}\BibitemShut {NoStop}%
\bibitem [{\citenamefont {Van~Vu}\ and\ \citenamefont
  {Hasegawa}(2021{\natexlab{a}})}]{Vu.2021.PRL}%
  \BibitemOpen
  \bibfield  {author} {\bibinfo {author} {\bibfnamefont {T.}~\bibnamefont
  {Van~Vu}}\ and\ \bibinfo {author} {\bibfnamefont {Y.}~\bibnamefont
  {Hasegawa}},\ }\bibfield  {title} {\bibinfo {title} {{Geometrical bounds of
  the irreversibility in Markovian systems}},\ }\href
  {https://doi.org/10.1103/PhysRevLett.126.010601} {\bibfield  {journal}
  {\bibinfo  {journal} {Phys. Rev. Lett.}\ }\textbf {\bibinfo {volume} {126}},\
  \bibinfo {pages} {010601} (\bibinfo {year} {2021}{\natexlab{a}})}\BibitemShut
  {NoStop}%
\bibitem [{\citenamefont {Maes}(2020)}]{Maes.2020.PR}%
  \BibitemOpen
  \bibfield  {author} {\bibinfo {author} {\bibfnamefont {C.}~\bibnamefont
  {Maes}},\ }\bibfield  {title} {\bibinfo {title} {{Frenesy: Time-symmetric
  dynamical activity in nonequilibria}},\ }\href
  {https://doi.org/10.1016/j.physrep.2020.01.002} {\bibfield  {journal}
  {\bibinfo  {journal} {Phys. Rep.}\ }\textbf {\bibinfo {volume} {850}},\
  \bibinfo {pages} {1} (\bibinfo {year} {2020})}\BibitemShut {NoStop}%
\bibitem [{Sup()}]{Supp.PhysRev}%
  \BibitemOpen
  \href@noop {} {}\bibinfo {note} {See Supplemental Material for detailed
  derivations of the main results and upper bounds of $\mathcal{Q}_1$ and
  $\mathcal{Q}_2$, which includes Ref.~\cite{Gammelmark.2014.PRL}.}\BibitemShut
  {Stop}%
\bibitem [{\citenamefont {van~den Bos}(2007)}]{Bos.2007}%
  \BibitemOpen
  \bibfield  {author} {\bibinfo {author} {\bibfnamefont {A.}~\bibnamefont
  {van~den Bos}},\ }\href@noop {} {\emph {\bibinfo {title} {{Parameter
  Estimation for Scientists and Engineers}}}}\ (\bibinfo  {publisher}
  {Wiley-Interscience},\ \bibinfo {address} {New York},\ \bibinfo {year}
  {2007})\BibitemShut {NoStop}%
\bibitem [{\citenamefont {Hiura}\ and\ \citenamefont
  {Sasa}(2021)}]{Hiura.2021.PRE}%
  \BibitemOpen
  \bibfield  {author} {\bibinfo {author} {\bibfnamefont {K.}~\bibnamefont
  {Hiura}}\ and\ \bibinfo {author} {\bibfnamefont {S.-i.}\ \bibnamefont
  {Sasa}},\ }\bibfield  {title} {\bibinfo {title} {{Kinetic uncertainty
  relation on first-passage time for accumulated current}},\ }\href
  {https://doi.org/10.1103/PhysRevE.103.L050103} {\bibfield  {journal}
  {\bibinfo  {journal} {Phys. Rev. E}\ }\textbf {\bibinfo {volume} {103}},\
  \bibinfo {pages} {L050103} (\bibinfo {year} {2021})}\BibitemShut {NoStop}%
\bibitem [{\citenamefont
  {Hasegawa}(2021{\natexlab{c}})}]{Hasegawa.2021.arxiv.QTURFPT}%
  \BibitemOpen
  \bibfield  {author} {\bibinfo {author} {\bibfnamefont {Y.}~\bibnamefont
  {Hasegawa}},\ }\bibfield  {title} {\bibinfo {title} {{Thermodynamic
  uncertainty relation for quantum first passage process via Loschmidt echo}},\
  }\href {https://arxiv.org/abs/2106.09870} {\bibfield  {journal} {\bibinfo
  {journal} {arXiv preprint arXiv:2106.09870}\ } (\bibinfo {year}
  {2021}{\natexlab{c}})}\BibitemShut {NoStop}%
\bibitem [{\citenamefont {Shiraishi}\ and\ \citenamefont
  {Saito}(2019)}]{Shiraishi.2019.PRL}%
  \BibitemOpen
  \bibfield  {author} {\bibinfo {author} {\bibfnamefont {N.}~\bibnamefont
  {Shiraishi}}\ and\ \bibinfo {author} {\bibfnamefont {K.}~\bibnamefont
  {Saito}},\ }\bibfield  {title} {\bibinfo {title} {{Information-theoretical
  bound of the irreversibility in thermal relaxation processes}},\ }\href
  {https://doi.org/10.1103/PhysRevLett.123.110603} {\bibfield  {journal}
  {\bibinfo  {journal} {Phys. Rev. Lett.}\ }\textbf {\bibinfo {volume} {123}},\
  \bibinfo {pages} {110603} (\bibinfo {year} {2019})}\BibitemShut {NoStop}%
\bibitem [{\citenamefont {Abiuso}\ \emph {et~al.}(2020)\citenamefont {Abiuso},
  \citenamefont {Miller}, \citenamefont {Perarnau-Llobet},\ and\ \citenamefont
  {Scandi}}]{Abiuso.2020.E}%
  \BibitemOpen
  \bibfield  {author} {\bibinfo {author} {\bibfnamefont {P.}~\bibnamefont
  {Abiuso}}, \bibinfo {author} {\bibfnamefont {H.~J.~D.}\ \bibnamefont
  {Miller}}, \bibinfo {author} {\bibfnamefont {M.}~\bibnamefont
  {Perarnau-Llobet}},\ and\ \bibinfo {author} {\bibfnamefont {M.}~\bibnamefont
  {Scandi}},\ }\bibfield  {title} {\bibinfo {title} {{Geometric optimisation of
  quantum thermodynamic processes}},\ }\href
  {https://doi.org/10.3390/e22101076} {\bibfield  {journal} {\bibinfo
  {journal} {Entropy}\ }\textbf {\bibinfo {volume} {22}},\ \bibinfo {pages}
  {1076} (\bibinfo {year} {2020})}\BibitemShut {NoStop}%
\bibitem [{\citenamefont {Van~Vu}\ and\ \citenamefont
  {Hasegawa}(2021{\natexlab{b}})}]{Vu.2021.PRL2}%
  \BibitemOpen
  \bibfield  {author} {\bibinfo {author} {\bibfnamefont {T.}~\bibnamefont
  {Van~Vu}}\ and\ \bibinfo {author} {\bibfnamefont {Y.}~\bibnamefont
  {Hasegawa}},\ }\bibfield  {title} {\bibinfo {title} {{Lower bound on
  irreversibility in thermal relaxation of open quantum systems}},\ }\href
  {https://doi.org/10.1103/PhysRevLett.127.190601} {\bibfield  {journal}
  {\bibinfo  {journal} {Phys. Rev. Lett.}\ }\textbf {\bibinfo {volume} {127}},\
  \bibinfo {pages} {190601} (\bibinfo {year} {2021}{\natexlab{b}})}\BibitemShut
  {NoStop}%
\bibitem [{\citenamefont {Geva}\ and\ \citenamefont
  {Kosloff}(1996)}]{Geva.1996.JCP}%
  \BibitemOpen
  \bibfield  {author} {\bibinfo {author} {\bibfnamefont {E.}~\bibnamefont
  {Geva}}\ and\ \bibinfo {author} {\bibfnamefont {R.}~\bibnamefont {Kosloff}},\
  }\bibfield  {title} {\bibinfo {title} {{The quantum heat engine and heat
  pump: An irreversible thermodynamic analysis of the three‐level
  amplifier}},\ }\href {https://doi.org/10.1063/1.471453} {\bibfield  {journal}
  {\bibinfo  {journal} {J. Chem. Phys.}\ }\textbf {\bibinfo {volume} {104}},\
  \bibinfo {pages} {7681} (\bibinfo {year} {1996})}\BibitemShut {NoStop}%
\bibitem [{\citenamefont {Boukobza}\ and\ \citenamefont
  {Tannor}(2007)}]{Boukobza.2007.PRL}%
  \BibitemOpen
  \bibfield  {author} {\bibinfo {author} {\bibfnamefont {E.}~\bibnamefont
  {Boukobza}}\ and\ \bibinfo {author} {\bibfnamefont {D.~J.}\ \bibnamefont
  {Tannor}},\ }\bibfield  {title} {\bibinfo {title} {{Three-level systems as
  amplifiers and attenuators: A thermodynamic analysis}},\ }\href
  {https://doi.org/10.1103/PhysRevLett.98.240601} {\bibfield  {journal}
  {\bibinfo  {journal} {Phys. Rev. Lett.}\ }\textbf {\bibinfo {volume} {98}},\
  \bibinfo {pages} {240601} (\bibinfo {year} {2007})}\BibitemShut {NoStop}%
\bibitem [{\citenamefont {Gammelmark}\ and\ \citenamefont
  {M\o{}lmer}(2014)}]{Gammelmark.2014.PRL}%
  \BibitemOpen
  \bibfield  {author} {\bibinfo {author} {\bibfnamefont {S.}~\bibnamefont
  {Gammelmark}}\ and\ \bibinfo {author} {\bibfnamefont {K.}~\bibnamefont
  {M\o{}lmer}},\ }\bibfield  {title} {\bibinfo {title} {{Fisher information and
  the quantum Cram\'er-Rao sensitivity limit of continuous measurements}},\
  }\href {https://doi.org/10.1103/PhysRevLett.112.170401} {\bibfield  {journal}
  {\bibinfo  {journal} {Phys. Rev. Lett.}\ }\textbf {\bibinfo {volume} {112}},\
  \bibinfo {pages} {170401} (\bibinfo {year} {2014})}\BibitemShut {NoStop}%
\end{thebibliography}
\end{document}


\title{Supplemental Material for \\ ``Thermodynamics of Precision in Markovian Open Quantum Dynamics''}

\author{Tan Van Vu}

\affiliation{Department of Physics, Keio University, 3-14-1 Hiyoshi, Kohoku-ku, Yokohama 223-8522, Japan}

\author{Keiji Saito}

\affiliation{Department of Physics, Keio University, 3-14-1 Hiyoshi, Kohoku-ku, Yokohama 223-8522, Japan}

\begin{abstract}
This supplemental material describes the details of the analytical calculations presented in the main text. 
The equations and figure numbers are prefixed with S [e.g., Eq.~(S1) or Fig.~S1]. 
The numbers without this prefix [e.g., Eq.~(1) or Fig.~1] refer to the items in the main text.
\end{abstract}

\pacs{}
\maketitle

\tableofcontents

\section{Lower bound on the irreversible entropy production rate}
Here, we show an analytical expression for the entropy production rate $\dot{\Sigma}_t$ and derive a lower bound on $\dot{\Sigma}_t$.
Taking the time derivative of the irreversible entropy production, the entropy production rate can be calculated as follows:
\begin{equation}
\begin{aligned}[b]
\dot{\Sigma}_t&=-\tr{\dot\varrho_t\ln\varrho_t}+\sum_k\tr{L_k^\dagger L_k\varrho_t}\Delta s_k\\
&=-\sum_k\tr{(\mca{D}[L_k]\varrho_t)\ln\varrho_t}+\sum_k\tr{L_k^\dagger L_k\varrho_t}\Delta s_k\\
&=\sum_k\tr{L_k\varrho_t(\Delta s_kL_k^\dagger-[L_k^\dagger,\ln\varrho_t])}.
\end{aligned}
\end{equation}
Let $\varrho_t=\sum_np_n(t)\dyad{n_t}$ be the spectral decomposition of the density matrix $\varrho_t$. We define $W_k^{nm}(t)\coloneqq|\mel{n_t}{L_k}{m_t}|^2$, which is always nonnegative.
Notice that $W_k^{nm}(t)=e^{\Delta s_k}W_{k'}^{mn}(t)$.
Because $\tr A=\sum_n\mel{n_t}{A}{n_t}$ for any operator $A$, the entropy production rate can be calculated as follows:
\begin{equation}\label{eq:ent.prod.rate}
\begin{aligned}[b]
\dot{\Sigma}_t&=\sum_k\sum_n\mel{n_t}{L_k\varrho_t(\Delta s_kL_k^\dagger-[L_k^\dagger,\ln\varrho_t])}{n_t}\\
&=\sum_k\sum_{n,m}W_k^{nm}(t)p_m(t)\qty[\Delta s_k+\ln\frac{p_m(t)}{p_n(t)}]\\
&=\frac{1}{2}\sum_k\sum_{n,m}W_k^{nm}(t)p_m(t)\qty[\Delta s_k+\ln\frac{p_m(t)}{p_n(t)}]+W_{k'}^{mn}(t)p_n(t)\qty[\Delta s_{k'}+\ln\frac{p_n(t)}{p_m(t)}]\\
&=\frac{1}{2}\sum_k\sum_{n,m}W_k^{nm}(t)p_m(t)\qty{\qty[\Delta s_k-\ln\frac{p_n(t)}{p_m(t)}]+e^{-\Delta s_k}\frac{p_n(t)}{p_m(t)}\qty[-\Delta s_k+\ln\frac{p_n(t)}{p_m(t)}]}\\
&=\frac{1}{2}\sum_k\sum_{n,m}W_k^{nm}(t)p_m(t)\qty[\Delta s_k-\ln\frac{p_n(t)}{p_m(t)}]\qty[1-e^{-\Delta s_k}\frac{p_n(t)}{p_m(t)}]\\
&=\frac{1}{2}\sum_k\sum_{n,m}\qty[W_k^{nm}(t)p_m(t)-W_{k'}^{mn}(t)p_n(t)]\ln\frac{W_k^{nm}(t)p_m(t)}{W_{k'}^{mn}(t)p_n(t)}.
\end{aligned}
\end{equation}
Because $(a-b)\ln(a/b)\ge 0$ for all $a,b\ge 0$, the positivity of $\dot{\Sigma}_t$ is immediately derived.

Notably, the following inequalities hold for arbitrary real numbers $a_n\ge 0,b_n\ge 0$:
\begin{align}
(a_1-b_1)\ln\frac{a_1}{b_1}&\ge 2\frac{(a_1-b_1)^2}{a_1+b_1},\\
\sum_n\frac{a_n^2}{b_n}&\ge\frac{\qty(\sum_na_n)^2}{\sum_nb_n}.
\end{align}
By applying the above inequalities to Eq.~\eqref{eq:ent.prod.rate}, we obtain a lower bound for the entropy production rate as follows:
\begin{equation}\label{eq:ent.prod.rate.lb}
\begin{aligned}[b]
\dot{\Sigma}_t&\ge\sum_k\sum_{n,m}\frac{\qty[W_k^{nm}(t)p_m(t)-W_{k'}^{mn}(t)p_n(t)]^2}{W_k^{nm}(t)p_m(t)+W_{k'}^{mn}(t)p_n(t)}\\
&\ge\sum_k\frac{\qty[\sum_{n,m}W_k^{nm}(t)p_m(t)-W_{k'}^{mn}(t)p_n(t)]^2}{\sum_{n,m}W_k^{nm}(t)p_m(t)+W_{k'}^{mn}(t)p_n(t)}\\
&=\sum_k\frac{\qty[\tr{L_k\varrho_tL_k^\dagger}-\tr{L_{k'}\varrho_tL_{k'}^\dagger}]^2}{\tr{L_k\varrho_tL_k^\dagger}+\tr{L_{k'}\varrho_tL_{k'}^\dagger}}.
\end{aligned}
\end{equation}
The lower bound in Eq.~\eqref{eq:ent.prod.rate.lb} will be used later to derive the quantum TUR for the currents.

\section{Derivation of the quantum TUR and KUR for dynamical and time observables}
Our derivation is based on the classical Cram{\'e}r-Rao inequality
\begin{equation}\label{eq:CR.ineq}
\frac{\Var{\Phi}}{(\partial_\theta\expval{\Phi}_\theta|_{\theta=0})^2}\ge\frac{1}{\mca{I}(0)},
\end{equation}
where the Fisher information is given by $\mca{I}(0)=-\expval{\partial_\theta^2\ln p_\theta(\Gamma_\tau)}|_{\theta=0}$.

In the following, we use $J$, $\Phi$, and $\Lambda$ to denote current-type, generic counting, and static observables, respectively.

\subsection{Derivation of the quantum TUR for currents}
We consider an auxiliary dynamics parameterized by the parameter $\theta$ as follows:
\begin{equation}\label{eq:modif.dyn.TUR}
H_\theta=(1+\theta)H,~L_{k,\theta}(t)=\sqrt{1+\ell_k(t)\theta}L_k,
\end{equation}
where
\begin{equation}
\ell_k(t)=\frac{\tr{L_{k}\varrho_tL_{k}^\dagger}-\tr{L_{k'}\varrho_tL_{k'}^\dagger}}{\tr{L_{k}\varrho_tL_{k}^\dagger}+\tr{L_{k'}\varrho_tL_{k'}^\dagger}}.
\end{equation}
The Hamiltonian and jump operators modified in Eq.~\eqref{eq:modif.dyn.TUR} are always valid for $\theta\ll 1$, and $\ell_k(t)=-\ell_{k'}(t)$.
Moreover, the auxiliary dynamics is reduced to the original when $\theta=0$.
The effective Hamiltonian and the non-unitary propagator have the following forms:
\begin{align}
H_{\rm eff,\theta}(t)&=(1+\theta)H-\frac{i}{2}\sum_k(1+\ell_k(t)\theta)L_k^\dagger L_k,\\
U_{\theta}(t_{j+1},t_j)&=\mca{T}\exp\qty{-i\int_{t_j}^{t_{j+1}}H_{\rm eff,\theta}(t)\dd{t}},
\end{align}
where $\mca{T}$ denotes the time-ordering operator.
The probability density of finding the trajectory $\Gamma_\tau$ in the auxiliary dynamics is given by
\begin{equation}
p_\theta(\Gamma_\tau)=p_n\prod_{j=1}^N(1+\ell_{k_j}(t_j)\theta)|U_{\theta}(\tau,t_N)\prod_{j=1}^{N}L_{k_{j}}U_{\theta}(t_{j},t_{j-1})\ket{n}|^2.
\end{equation}
Subsequently, the Fisher information can be calculated as
\begin{equation}\label{eq:FI.lb.TUR}
\begin{aligned}[b]
\mca{I}(0)&=-\eval{\expval{\partial_\theta^2\ln\prod_{j=1}^N(1+\ell_{k_j}(t_j)\theta)}}_{\theta=0} - \eval{\expval{\partial_\theta^2\ln|U_{\theta}(\tau,t_N)\prod_{j=1}^{N}L_{k_{j}}U_{\theta}(t_{j},t_{j-1})\ket{n}|^2}}_{\theta=0}\\
&=\expval{\sum_{j=1}^N\ell_{k_j}(t_j)^2} +\expval{q_1(\Gamma_\tau)}\\
&=\int_0^\tau\sum_k \tr{L_k\varrho_tL_k^\dagger}\ell_k(t)^2\dd{t} +\expval{q_1(\Gamma_\tau)}\\
&=\frac{1}{2}\int_0^\tau\sum_k\qty{\tr{L_k\varrho_tL_k^\dagger}+\tr{L_{k'}\varrho_tL_{k'}^\dagger}}\ell_k(t)^2\dd{t} +\expval{q_1(\Gamma_\tau)}\\
&=\frac{1}{2}\int_0^\tau\sum_k\frac{\qty[\tr{L_{k}\varrho_tL_{k}^\dagger}-\tr{L_{k'}\varrho_tL_{k'}^\dagger}]^2}{\tr{L_{k}\varrho_tL_{k}^\dagger}+\tr{L_{k'}\varrho_tL_{k'}^\dagger}}\dd{t} +\expval{q_1(\Gamma_\tau)}\\
&\le\frac{1}{2}\int_0^\tau\dot{\Sigma}_t\dd{t} +\expval{q_1(\Gamma_\tau)}\\
&=\Sigma_\tau/2+\mca{Q}_1.
\end{aligned}
\end{equation}
Here, we have used Eq.~\eqref{eq:ent.prod.rate.lb} to obtain the last inequality and have defined
\begin{align}
\mca{Q}_1&\coloneqq \expval{q_1(\Gamma_\tau)},\\
q_1(\Gamma_\tau)&\coloneqq-\eval{\partial_\theta^2\ln|U_{\theta}(\tau,t_N)\prod_{j=1}^{N}L_{k_{j}}U_{\theta}(t_{j},t_{j-1})\ket{n}|^2}_{\theta=0}.
\end{align}

For $\theta\ll 1$, the density operator $\varrho_{t,\theta}$ in the auxiliary dynamics can be expanded in terms of $\theta$ as $\varrho_{t,\theta}=\varrho_t+\theta\phi_t+\order{\theta^2}$.
Substituting this perturbative expression to the Lindblad master equation, we have
\begin{equation}
\dot\varrho_t+\theta\dot\phi_t=-i[(1+\theta)H,\varrho_t+\theta\phi_t]+\sum_k(1+\ell_k(t)\theta)\qty[L_k(\varrho_t+\theta\phi_t)L_k^\dagger-\frac{1}{2}\{L_k^\dagger L_k,\varrho_t+\theta\phi_t\}]+\order{\theta^2}.
\end{equation}
By collecting the terms in the first order of $\theta$, we obtain the differential equation describing the time evolution of the operator $\phi_t$:
\begin{equation}
\dot\phi_t=-i[H,\varrho_t+\phi_t]+\sum_k\qty{\mca{D}[L_k]\phi_t+\ell_k(t)\mca{D}[L_k]\varrho_t}=\mca{L}(\varrho_t+\phi_t)+\sum_k[\ell_k(t)-1]\mca{D}[L_k]\varrho_t
\end{equation}
given the initial condition $\phi_0=\mbb{0}$.
It can be easily seen that the operator $\phi_t$ is always traceless.
Noting that $w_k=-w_{k'}$ and $w_k\ell_k(t)=w_{k'}\ell_{k'}(t)$, the partial derivative of the current average in the auxiliary dynamics with respect to $\theta$ can be calculated as
\begin{equation}\label{eq:par.avg.TUR}
\begin{aligned}[b]
\eval{\partial_\theta\expval{J}_\theta}_{\theta=0}&=\eval{\partial_\theta\qty[\int_0^\tau\sum_kw_k(1+\ell_k(t)\theta)\tr{L_k(\varrho_t+\theta\phi_t)L_k^\dagger}\dd{t}+O(\theta^2)]}_{\theta=0}\\
&=\int_0^\tau\sum_kw_k\ell_k(t)\tr{L_k\varrho_tL_k^\dagger}\dd{t}+\int_0^\tau\sum_kw_k\tr{L_k\phi_tL_k^\dagger}\dd{t}\\
&=\frac{1}{2}\int_0^\tau\sum_kw_k\ell_k(t)\qty[\tr{L_k\varrho_tL_k^\dagger}+\tr{L_{k'}\varrho_tL_{k'}^\dagger}]\dd{t}+\int_0^\tau\sum_kw_k\tr{L_k\phi_tL_k^\dagger}\dd{t}\\
&=\frac{1}{2}\int_0^\tau\sum_kw_k\qty[\tr{L_k\varrho_tL_k^\dagger}-\tr{L_{k'}\varrho_tL_{k'}^\dagger}]\dd{t}+\int_0^\tau\sum_kw_k\tr{L_k\phi_tL_k^\dagger}\dd{t}\\
&=\int_0^\tau\sum_kw_k\tr{L_k\varrho_tL_k^\dagger}\dd{t}+\int_0^\tau\sum_kw_k\tr{L_k\phi_tL_k^\dagger}\dd{t}\\
&=\expval{J}+\expval{J}_*,
\end{aligned}
\end{equation}
where we have defined $\expval{J}_*\coloneqq\int_0^\tau\sum_kw_k\tr{L_k\phi_tL_k^\dagger}\dd{t}$.
From Eqs.~\eqref{eq:FI.lb.TUR} and \eqref{eq:par.avg.TUR}, we obtain the quantum TUR,
\begin{equation}
\frac{\Var{J}}{\expval{J}^2}\ge\frac{2(1+\tilde{\delta} J)^2}{\Sigma_\tau+2\mca{Q}_1},
\end{equation}
where $\tilde{\delta}J\coloneqq\expval{J}_*/\expval{J}$.

\subsection{Derivation of the quantum KUR for counting observables}

We consider auxiliary dynamics with the Hamiltonian and jump operators are parameterized by the parameter $\theta$ as follows:
\begin{equation}
H_\theta=(1+\theta)H,~L_{k,\theta}=\sqrt{1+\theta}L_k.
\end{equation}
Evidently, the auxiliary dynamics is reduced to the original when $\theta=0$.
The effective Hamiltonian and the non-unitary propagator have the following forms:
\begin{align}
H_{\rm eff,\theta}&=(1+\theta)H-\frac{i}{2}\sum_k(1+\theta)L_k^\dagger L_k=(1+\theta)H_{\rm eff},\\
U_{\theta}(t_{j+1},t_j)&=\exp\qty[-iH_{\rm eff,\theta}(t_{j+1}-t_j)].
\end{align}
The probability density of observing the trajectory $\Gamma_\tau$ is given by
\begin{equation}
p_\theta(\Gamma_\tau)=p_n(1+\theta)^N|U_{\theta}(\tau,t_N)\prod_{j=1}^{N}L_{k_{j}}U_{\theta}(t_{j},t_{j-1})\ket{n}|^2.
\end{equation}
Analogously, the Fisher information can be calculated as
\begin{equation}\label{eq:FI.lb.KUR}
\begin{aligned}[b]
\mca{I}(0)&=-\eval{\expval{\partial_\theta^2\ln(1+\theta)^N}}_{\theta=0} - \eval{\expval{\partial_\theta^2\ln|U_{\theta}(\tau,t_N)\prod_{j=1}^{N}L_{k_{j}}U_{\theta}(t_{j},t_{j-1})\ket{n}|^2}}_{\theta=0}\\
&=\expval{N} +\expval{q_2(\Gamma_\tau)}\\
&=\int_0^\tau\tr{L_k\varrho_tL_k^\dagger}\dd{t} +\expval{q_2(\Gamma_\tau)}\\
&=\mca{A}_\tau+\mca{Q}_2.
\end{aligned}
\end{equation}
Here, we have defined
\begin{align}
\mca{Q}_2&\coloneqq \expval{q_2(\Gamma_\tau)},\\
q_2(\Gamma_\tau)&\coloneqq-\eval{\partial_\theta^2\ln|U_{\theta}(\tau,t_N)\prod_{j=1}^{N}L_{k_{j}}U_{\theta}(t_{j},t_{j-1})\ket{n}|^2}_{\theta=0}.
\end{align}
Note that the quantities $q_1$ and $q_2$ are not the same because the form of the propagator $U_{\theta}$ is different in the two cases.

It can be verified that the density operator in the auxiliary dynamics is related to that in the original as $\varrho_{t,\theta}=\varrho_{t(1+\theta)}$.
Therefore, the partial derivative of the average of the generic observable in the auxiliary dynamics with respect to $\theta$ can be calculated as
\begin{equation}\label{eq:par.avg.KUR}
\begin{aligned}[b]
\eval{\partial_\theta\expval{\Phi}_\theta}_{\theta=0}&=\eval{\partial_\theta\qty[\int_0^\tau\sum_kw_k(1+\theta)\tr{L_k\varrho_{t,\theta}L_k^\dagger}\dd{t}]}_{\theta=0}\\
&=\eval{\partial_\theta\qty[\int_0^\tau\sum_kw_k(1+\theta)\tr{L_k\varrho_{t(1+\theta)}L_k^\dagger}\dd{t}]}_{\theta=0}\\
&=\eval{\partial_\theta\qty[\int_0^{(1+\theta)\tau}\sum_kw_k\tr{L_k\varrho_{t}L_k^\dagger}\dd{t}]}_{\theta=0}\\
&=\tau\sum_kw_k\tr{L_k\varrho_{\tau}L_k^\dagger}\\
&=\tau\partial_\tau\expval{\Phi}.
\end{aligned}
\end{equation}
From Eqs.~\eqref{eq:FI.lb.KUR} and \eqref{eq:par.avg.KUR}, we obtain the quantum KUR as follows:
\begin{equation}
\frac{\Var{\Phi}}{\expval{\Phi}^2}\ge\frac{(1+\delta\Phi)^2}{\mca{A}_\tau+\mca{Q}_2},
\end{equation}
where $\delta\Phi\coloneqq\tau\partial_\tau\ln|\expval{\Phi}/\tau|$.

\subsection{Derivation of the quantum KUR for static observables}
We consider the same auxiliary dynamics as in the previous section:
\begin{equation}\label{eq:modif.dyn.KUR}
H_\theta=(1+\theta)H,~L_{k,\theta}(t)=\sqrt{1+\theta}L_k.
\end{equation}
The Fisher information can be analogously calculated using Eq.~\eqref{eq:FI.lb.KUR}.
Because $\varrho_{t,\theta}=\varrho_{t(1+\theta)}$, the partial derivative of the average of the static observable in the auxiliary dynamics with respect to $\theta$ can be calculated as
\begin{equation}
\begin{aligned}[b]
\eval{\partial_\theta\expval{\Lambda}_\theta}_{\theta=0}&=\eval{\partial_\theta\qty[\tau^{-1}\int_0^\tau\tr{A\varrho_{t,\theta}}\dd{t}]}_{\theta=0}\\
&=\eval{\partial_\theta\qty[\tau^{-1}\int_0^\tau\tr{A\varrho_{t(1+\theta)}}\dd{t}]}_{\theta=0}\\
&=\eval{\partial_\theta\qty[(1+\theta)^{-1}\tau^{-1}\int_0^{(1+\theta)\tau}\tr{A\varrho_{t}}\dd{t}]}_{\theta=0}\\
&=-\expval{\Lambda}+\tr{A\varrho_{\tau}}\\
&=-\expval{\Lambda}+\partial_\tau(\tau\expval{\Lambda})\\
&=\tau\partial_\tau\expval{\Lambda}.
\end{aligned}
\end{equation}
Consequently, we obtain the quantum KUR for the static observable as the following:
\begin{equation}
\frac{\Var{\Lambda}}{\expval{\Lambda}^2}\ge\frac{(1+\delta\Lambda)^2}{\mca{A}_\tau+\mca{Q}_2},
\end{equation}
where $\delta\Lambda\coloneqq\tau\partial_\tau\ln|\expval{\Lambda}/\tau|$.

\subsection{Derivation of the quantum KUR for the first passage time}

We consider the same auxiliary dynamics as in the derivation of the quantum KUR, that is, the Hamiltonian and jump operators are parameterized as follows:
\begin{equation}
H_\theta=(1+\theta)H,~L_{k,\theta}=\sqrt{1+\theta}L_k.
\end{equation}
We assume that $P(\tau<+\infty)=1$ and the mean and variance of $\tau$ are finite.
Applying the Cram{\'e}r-Rao inequality, we have
\begin{equation}
\frac{\Var{\tau}}{(\partial_\theta\expval{\tau}_\theta|_{\theta=0})^2}\ge\frac{1}{\mca{I}(0)}.
\end{equation}
For the first passage time problem, the time at which the last jump occurs is always the stopping time. That is, $\tau=t_N$, where $N=\min\{m\,|\,\sum_{j=1}^mw_{k_j}\ge\Phi_{\rm thr}\}$.
Each trajectory $\Gamma_\tau$ can be described by the discrete set $\{(t_0,n),(t_1,k_1),\dots,(t_N,k_N)\}$.
Let $\tilde\Gamma_{\rm jump}=\{k_1,\dots,k_N\}$ be the trajectory of only jump events, then the partial derivative of the average of the first passage time in the auxiliary dynamics with respect to $\theta$ can be calculated as follows:
\begin{equation}
\begin{aligned}[b]
\partial_\theta\expval{\tau}_\theta|_{\theta=0}&=\eval{\partial_\theta\qty[\sum_{\tilde\Gamma_{\rm jump}}\int_0^\infty\dd{\tau}\int_{0}^{\tau}\dd{t_1}\int_{t_1}^{\tau}\dd{t_2}~\dots\int_{t_{N-2}}^{\tau}\dd{t_{N-1}}p_n(1+\theta)^N|\prod_{j=1}^{N}L_{k_{j}}U_{\theta}(t_{j},t_{j-1})\ket{n}|^2\tau]}_{\theta=0}\\
&=\eval{\partial_\theta\qty[\sum_{\tilde\Gamma_{\rm jump}}\int_0^\infty\dd{\tau'}\int_{0}^{\tau'}\dd{t_1'}\int_{t_1'}^{\tau'}\dd{t_2'}~\dots\int_{t_{N-2}'}^{\tau'}\dd{t_{N-1}'}p_n|\prod_{j=1}^{N}L_{k_{j}}U(t_{j}',t_{j-1}')\ket{n}|^2(1+\theta)^{-1}\tau']}_{\theta=0}\\
&=\partial_\theta[(1+\theta)^{-1}\expval{\tau}]|_{\theta=0}\\
&=-\expval{\tau}.
\end{aligned}
\end{equation}
Here we have changed the variables $t_j'=(1+\theta)t_j$ and $\tau'=(1+\theta)\tau$ in the integration.
Likewise, the Fisher information can be calculated as
\begin{equation}
\begin{aligned}[b]
\mca{I}(0)&=-\left.\expval{\partial_\theta^2\ln(1+\theta)^N}\right|_{\theta=0} - \left.\expval{\partial_\theta^2\ln|\prod_{j=1}^{N}L_{k_{j}}U_{\theta}(t_{j},t_{j-1})\ket{n}|^2}\right|_{\theta=0}\\
&=\expval{N}_\tau +\expval{q_3(\Gamma_\tau)}\\
&=\expval{N}_\tau +\mca{Q}_3,
\end{aligned}
\end{equation}
where we have defined
\begin{align}
\mca{Q}_3&\coloneqq \expval{q_3(\Gamma_\tau)},\\
q_3(\Gamma_\tau)&\coloneqq-\left.\partial_\theta^2\ln|\prod_{j=1}^{N}L_{k_{j}}U_{\theta}(t_{j},t_{j-1})\ket{n}|^2\right|_{\theta=0}.
\end{align}
Consequently, we obtain the quantum KUR for the first passage time as follows:
\begin{equation}
\frac{\Var{\tau}}{\expval{\tau}^2}\ge\frac{1}{\expval{N}_\tau+\mca{Q}_3}.
\end{equation}

In the above, we have assumed that $P(\tau<+\infty)=1$. Nevertheless, an analogous bound can be derived in the remaining case [i.e., when the probability of an infinite stopping time is positive, $P(\tau<+\infty)<1$]. To this end, we define the average for arbitrary functional function $f$ as
\begin{equation}
\expval{f}_{\bullet,\theta}\coloneqq\sumint f(\Gamma_\tau) p_{\theta}(\Gamma_\tau)\dd{\Gamma_\tau},
\end{equation}
where the average is over all trajectories with finite stopping times. For simplicity, we denote $\expval{\cdot}_{\bullet,\theta}$ by $\expval{\cdot}_\bullet$ for the $\theta=0$ case. The modified variance of the stopping time can be defined as
\begin{equation}
\Var{\tau}_\bullet\coloneqq \expval{\qty(\tau-\expval{\tau}_\bullet)^2}_\bullet.
\end{equation}
When $P(\tau<+\infty)=1$, these modified mean and variance reduce to $\expval{\tau}$ and $\Var{\tau}$, respectively.
Note that $\partial_\theta\expval{1}_{\bullet,\theta}=0$.
The same approach using the Cram{\'e}r-Rao inequality yields the following KUR:
\begin{equation}
\frac{\Var{\tau}_\bullet}{\expval{\tau}_\bullet^2}\ge\frac{1}{\expval{N}_{\tau,\bullet}+\expval{q_3}_\bullet}.
\end{equation}

\section{Upper bounds of $\mca{Q}_1$ and $\mca{Q}_2$ in the long-time regime}

Here we show that in the long-time regime, the terms $\mca{Q}_1$ and $\mca{Q}_2$ can be upper bounded by simple quantities that depend only on the Hamiltonian, jump operators, and the stationary density matrix.
Notice that the obtained Fisher information is always upper bounded by the quantum Fisher information, which is maximized over all positive operator valued measures (POVMs),
\begin{equation}
\mca{I}(0)\le \mca{I}_Q\coloneqq \max_{\mca{P}}\qty{\mca{I}(0;\mca{P})}.
\end{equation}
Here, $\mca{I}(0;\mca{P})$ is the Fisher information obtained using a specific POVM $\mca{P}$.
Moreover, for long-time steady-state systems, the quantum Fisher information can be explicitly calculated \cite{Gammelmark.2014.PRL},
\begin{equation}
\mca{I}_Q=4\tau\eval{\partial^2_{\theta_1\theta_2}\chi(\vb*{\theta})}_{\vb*{\theta}=\vb*{0}},
\end{equation}
where $\chi(\vb*{\theta})~(\vb*{\theta}=[ \theta_1,\theta_2 ]^\top)$ is the dominant eigenvalue of the generalized Lindblad super-operator
\begin{equation}
\mca{L}_{\vb*{\theta}}(\varrho)=-i[(1+\theta_1)H\varrho-(1+\theta_2)\varrho H]+\sum_k\sqrt{(1+\ell_k\theta_1)(1+\ell_k\theta_2)}L_k\varrho L_k^\dagger-\frac{1}{2}\sum_k(1+\ell_k\theta_1)L_k^\dagger L_k\varrho-\frac{1}{2}\sum_k(1+\ell_k\theta_2)\varrho L_k^\dagger L_k.
\end{equation}
To calculate $\mca{I}_Q$, it is convenient to vectorize operators as
\begin{equation}
X=\sum_{m,n}x_{mn}\dyad{m}{n}\rightarrow \kvec{X}=\sum_{m,n}x_{mn}\ket{m}\otimes\ket{n}.
\end{equation}
We can easily show that $\kvec{XY}=(X\otimes\mbb{1})\kvec{Y}$ and $\kvec{YX}=(\mbb{1}\otimes X^\top)\kvec{Y}$.
Using this representation, the Lindblad equation can be written as
\begin{equation}
\kvec{\dot\varrho_t}=\hat{\mca{L}}\kvec{\varrho_t},
\end{equation}
where the operator $\hat{\mca{L}}$ is defined as
\begin{equation}
\hat{\mca{L}}=-i(H\otimes\mbb{1} - \mbb{1}\otimes H^\top)+\sum_k\qty[ L_k\otimes L_k^* - \frac{1}{2} ( L_k^\dagger L_k )\otimes\mbb{1} - \frac{1}{2} \mbb{1}\otimes( L_k^\dagger L_k )^\top].
\end{equation}
Here $\top$ and $*$ denote the matrix transpose and complex conjugate, respectively.
For the matrix $\hat{\mca{L}}_{\vb*{\theta}}$, let $\vb*{u}(\vb*{\theta})$ and $\vb*{v}(\vb*{\theta})$ denote the corresponding left and right eigenvectors associated with an eigenvalue $\chi(\vb*{\theta})$ (which is vanished at $\vb*{\theta}=\vb*{0}$),
\begin{align}
\hat{\mca{L}}_{\vb*{\theta}}\vb*{u}(\vb*{\theta})&=\chi(\vb*{\theta})\vb*{u}(\vb*{\theta}),\\
\hat{\mca{L}}_{\vb*{\theta}}^\dagger \vb*{v}(\vb*{\theta})&=\chi(\vb*{\theta})^*\vb*{v}(\vb*{\theta}).
\end{align}
Here, $\vb*{u}(\vb*{\theta})$ and $\vb*{v}(\vb*{\theta})$ satisfy the normalization constraints, $\expval{\vb*{u}(\vb*{0}),\vb*{u}(\vb*{\theta})}=1$ and $\expval{\vb*{u}(\vb*{0}),\vb*{v}(\vb*{0})}=1$, where we have used the notation of the Frobenius inner product $\expval{X,Y}=\tr{X^\dagger Y}$.
Specifically, 
\begin{align}
\vb*{u}(\vb*{0})=\kvec{\varrho^{\rm ss}}/\sqrt{\bkvec{\varrho^{\rm ss}}},\\
\vb*{v}(\vb*{0})=\kvec{\mbb{1}}\sqrt{\bkvec{\varrho^{\rm ss}}}.
\end{align}
Here, $\varrho^{\rm ss}$ denotes the steady-state density matrix.
Then, the partial derivative of the eigenvalue $\chi(\vb*{\theta})$ can be calculated as
\begin{equation}\label{eq:eigen.deriv.1}
\eval{\partial^2_{\theta_1\theta_2}\chi(\vb*{\theta})}_{\vb*{\theta}=\vb*{0}}=\expval{\vb*{v}(\vb*{0}),\partial^2_{\theta_1\theta_2}\hat{\mca{L}}_{\vb*{\theta}} \vb*{u}(\vb*{0})}_{\vb*{\theta}=\vb*{0}}-\expval{\vb*{v}(\vb*{0}),\partial_{\theta_1}\hat{\mca{L}}_{\vb*{\theta}}\mbb{P}\hat{\mca{L}}^{+}\mbb{P}\partial_{\theta_2}\hat{\mca{L}}_{\vb*{\theta}}\vb*{u}(\vb*{0})+\partial_{\theta_2}\hat{\mca{L}}_{\vb*{\theta}}\mbb{P}\hat{\mca{L}}^{+}\mbb{P}\partial_{\theta_1}\hat{\mca{L}}_{\vb*{\theta}}\vb*{u}(\vb*{0})}_{\vb*{\theta}=\vb*{0}}.
\end{equation}
Here, the operator $\mbb{P}$ denotes the projection onto the complement of the $0$-eigenspace (i.e., $\mbb{P}\vb*{x}=\vb*{x}-\expval{\vb*{v}(\vb*{0}),\vb*{x}}\vb*{u}(\vb*{0})$) and $\hat{\mca{L}}^+$ denotes the Moore-Penrose pseudo-inverse of $\hat{\mca{L}}$.
The first term on the right-hand side of Eq.~\eqref{eq:eigen.deriv.1} can be explicitly calculated as
\begin{equation}
\expval{\vb*{v}(\vb*{0}),\partial^2_{\theta_1\theta_2}\hat{\mca{L}}_{\vb*{\theta}} \vb*{u}(\vb*{0})}_{\vb*{\theta}=\vb*{0}}=\frac{1}{4}\bvec{\mbb{1}}\sum_k\ell_k^2(L_k\otimes L_k^*)\kvec{\varrho^{\rm ss}}=\frac{1}{4}\sum_k\ell_k^2\tr{L_k\varrho^{\rm ss}L_k^\dagger}.
\end{equation}
The second term can be analogously calculated as
\begin{align}
&\expval{\vb*{v}(\vb*{0}),\partial_{\theta_1}\hat{\mca{L}}_{\vb*{\theta}}\mbb{P}\hat{\mca{L}}^{+}\mbb{P}\partial_{\theta_2}\hat{\mca{L}}_{\vb*{\theta}}\vb*{u}(\vb*{0})+\partial_{\theta_2}\hat{\mca{L}}_{\vb*{\theta}}\mbb{P}\hat{\mca{L}}^{+}\mbb{P}\partial_{\theta_1}\hat{\mca{L}}_{\vb*{\theta}}\vb*{u}(\vb*{0})}_{\vb*{\theta}=\vb*{0}}\\
&=\bvec{\mbb{1}}\mca{F}_1\mbb{P}\hat{\mca{L}}^{+}\mbb{P}\mca{F}_2\kvec{\varrho^{\rm ss}}+\bvec{\mbb{1}}\mca{F}_2\mbb{P}\hat{\mca{L}}^{+}\mbb{P}\mca{F}_1\kvec{\varrho^{\rm ss}},
\end{align}
where the matrices $\mca{F}_1$ and $\mca{F}_2$ are given by
\begin{align}
\mca{F}_1&\coloneqq -iH\otimes\mbb{1}+\frac{1}{2}\sum_k\ell_k\qty[ L_k\otimes L_k^* - (L_k^\dagger L_k )\otimes\mbb{1}],\\
\mca{F}_2&\coloneqq i\mbb{1}\otimes H^\top+\frac{1}{2}\sum_k\ell_k\qty[ L_k\otimes L_k^* - \mbb{1}\otimes(L_k^\dagger L_k )^\top].
\end{align}
Define ${\mca{Q}}_1^{u}\coloneqq -4\tau \qty( \bvec{\mbb{1}}\mca{F}_1\mbb{P}\hat{\mca{L}}^{+}\mbb{P}\mca{F}_2\kvec{\varrho^{\rm ss}}+\bvec{\mbb{1}}\mca{F}_2\mbb{P}\hat{\mca{L}}^{+}\mbb{P}\mca{F}_1\kvec{\varrho^{\rm ss}} )$, we readily obtain
\begin{equation}
\mca{I}_Q=\tau\sum_k\ell_k^2\tr{L_k\varrho^{\rm ss}L_k^\dagger}+{\mca{Q}}_1^{u}.
\end{equation}
Since $\mca{I}(0)\le \mca{I}_Q$ and
\begin{equation}
\mca{I}(0)=\tau\sum_k\ell_k^2\tr{L_k\varrho^{\rm ss}L_k^\dagger}+\mca{Q}_1,
\end{equation}
we consequently obtain the following upper bound for $\mca{Q}_1$:
\begin{equation}
\mca{Q}_1\le {\mca{Q}}_1^{u}=-4\tau \qty( \bvec{\mbb{1}}\mca{F}_1\mbb{P}\hat{\mca{L}}^{+}\mbb{P}\mca{F}_2\kvec{\varrho^{\rm ss}}+\bvec{\mbb{1}}\mca{F}_2\mbb{P}\hat{\mca{L}}^{+}\mbb{P}\mca{F}_1\kvec{\varrho^{\rm ss}} ).
\end{equation}
Unlike the term $\mca{Q}_1$ (which requires information of all trajectories to calculate), the term ${\mca{Q}}_1^{u}$ has a simpler form and can be calculated using only the Hamiltonian, jump operators, and the steady-state density matrix.
\begin{figure}[b]
\centering
\includegraphics[width=0.6\linewidth]{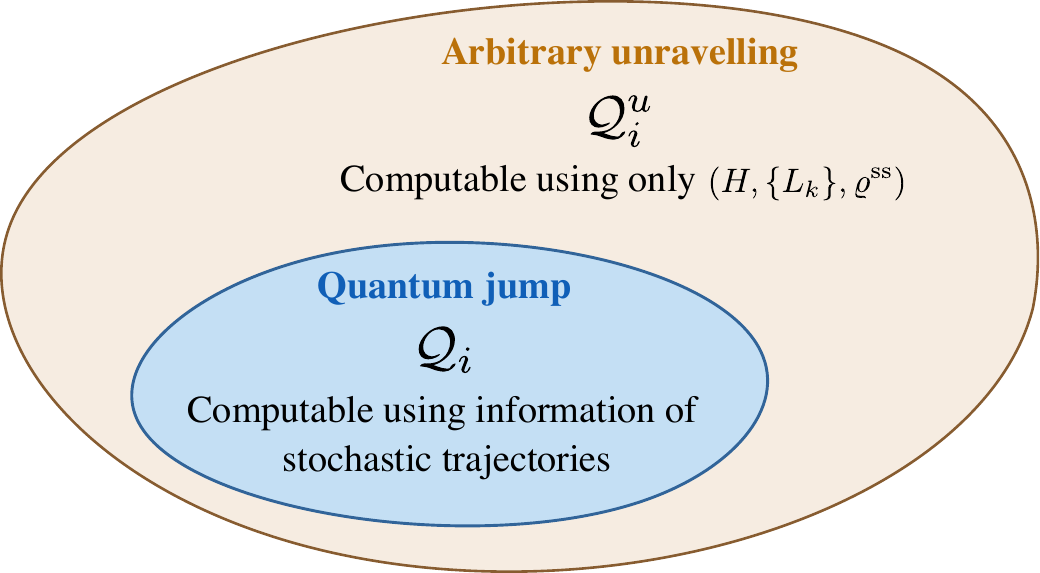}
\protect\caption{The relationship between $\mca{Q}_i$ and its upper bound $\mca{Q}_i^u$. While calculating $\mca{Q}_i$ requires information of stochastic trajectories, its upper bound $\mca{Q}_i^u$ can be easily calculated using solely the Hamiltonian $H$, jump operators $\{L_k\}$, and the steady-state density matrix $\varrho^{\rm ss}$. The upper bound $\mca{Q}_i^u$ is the quantum contribution that can be applied to an arbitrary unraveling of the Lindblad master equation.}\label{fig:Qi}
\end{figure}
The relationship between $\mca{Q}_1$ and $\mca{Q}_1^u$ is illustrated in Fig.~\ref{fig:Qi}.
Consequently, we obtain the following hierarchy of lower bounds for the fluctuation of currents:
\begin{equation}
\frac{\Var{J}}{\expval{J}^2}\ge\frac{2(1+\tilde{\delta} J)^2}{\Sigma_\tau+2\mca{Q}_1}\ge \frac{2(1+\tilde{\delta} J)^2}{\Sigma_\tau+2{\mca{Q}}_1^{u}}.
\end{equation}
In the following, we show that $\mca{Q}_1^u=0$ in the classical limit (i.e., $H=\mbb{0}$ and $L_k=\sqrt{\gamma_{mn}}\dyad{\epsilon_m}{\epsilon_n}$).
Nevertheless, it should be noted that $\mca{Q}_1^u$ does not vanish in the general case; therefore, $\mca{Q}_1^u$ is identified as a quantum term.

Let $p_n=\mel{\epsilon_n}{\varrho^{\rm ss}}{\epsilon_n}$, then the steady-state condition gives
\begin{equation}
\begin{aligned}[b]
-i[H, \varrho^{\rm ss} ] + \sum_{k}\qty( L_k\varrho^{\rm ss}L_k^\dagger - \frac{1}{2}\qty{ L_k^\dagger L_k,\varrho^{\rm ss} } ) = 0 &\to \sum_{m\neq n}\qty(\gamma_{mn}p_n\dyad{\epsilon_m}{\epsilon_m} - \gamma_{mn}p_n\dyad{\epsilon_n}{\epsilon_n}) = 0\\
&\to \sum_m\sum_{n(\neq m)}\qty(\gamma_{mn}p_n-\gamma_{nm}p_m)\dyad{\epsilon_m}{\epsilon_m}=0.
\end{aligned}
\end{equation}
By using $\ell_k=(\gamma_{mn}p_n-\gamma_{nm}p_m)/(\gamma_{mn}p_n+\gamma_{nm}p_m)$, we can calculate as follows:
\begin{equation}
\begin{aligned}[b]
\mca{F}_1\kvec{\varrho^{\rm ss}}&=\kvec{-iH\varrho^{\rm ss}+\frac{1}{2}\sum_k\ell_k\qty(L_k\varrho^{\rm ss}L_k^\dagger-L_k^\dagger L_k\varrho^{\rm ss})}\\
&=\frac{1}{2}\kvec{\sum_{m\neq n}\frac{\gamma_{mn}p_n-\gamma_{nm}p_m}{\gamma_{mn}p_n+\gamma_{nm}p_m}\qty( \gamma_{mn}p_n\dyad{\epsilon_m}{\epsilon_m} -\gamma_{mn}p_n\dyad{\epsilon_n}{\epsilon_n} ) }\\
&=\frac{1}{2}\kvec{\sum_{m}\sum_{n(\neq m)}\frac{\gamma_{mn}p_n-\gamma_{nm}p_m}{\gamma_{mn}p_n+\gamma_{nm}p_m}\qty( \gamma_{mn}p_n + \gamma_{nm}p_m)\dyad{\epsilon_m}{\epsilon_m}}\\
&=\frac{1}{2}\kvec{\sum_{m}\sum_{n(\neq m)}\qty( \gamma_{mn}p_n - \gamma_{nm}p_m)\dyad{\epsilon_m}{\epsilon_m}}\\
&=0.
\end{aligned}
\end{equation}
Analogously, we can also show that $\mca{F}_2\kvec{\varrho^{\rm ss}}=0$.
Consequently, we obtain $\mca{Q}_1^u=0$.

Following the same procedure, we also obtain the following upper bound for $\mca{Q}_2$:
\begin{equation}
\mca{Q}_2\le{\mca{Q}}_2^{u}\coloneqq -4\tau \qty( \bvec{\mbb{1}}\mca{G}_1\mbb{P}\hat{\mca{L}}^{+}\mbb{P}\mca{G}_2\kvec{\varrho^{\rm ss}}+\bvec{\mbb{1}}\mca{G}_2\mbb{P}\hat{\mca{L}}^{+}\mbb{P}\mca{G}_1\kvec{\varrho^{\rm ss}} ),
\end{equation}
where the matrices $\mca{G}_1$ and $\mca{G}_2$ are given by
\begin{align}
\mca{G}_1&\coloneqq -iH\otimes\mbb{1}+\frac{1}{2}\sum_k\qty[ L_k\otimes L_k^* - (L_k^\dagger L_k )\otimes\mbb{1}],\\
\mca{G}_2&\coloneqq i\mbb{1}\otimes H^\top+\frac{1}{2}\sum_k\qty[ L_k\otimes L_k^* - \mbb{1}\otimes(L_k^\dagger L_k )^\top].
\end{align}
Note that the term $\mca{Q}_2^u$ also vanishes in the classical limit.

\subsection{Explicit expressions of the upper bounds in a two-level system}
Here we explicitly calculate the upper bounds $\mca{Q}_1^u$ and $\mca{Q}_2^u$ for a two-level system.
Consider an open quantum system with the following Hamiltonian and jump operators:
\begin{align}
H&=\Delta\dyad{\epsilon_1}{\epsilon_1}+\Omega( \dyad{\epsilon_1}{\epsilon_0} + \dyad{\epsilon_0}{\epsilon_1} ),\\
L_1&=\sqrt{\gamma n}\dyad{\epsilon_1}{\epsilon_0},\\
L_{1'}&=\sqrt{\gamma (n + 1)}\dyad{\epsilon_0}{\epsilon_1}.
\end{align}
Here, $\Delta$ and $\Omega$ are real parameters, and $\ket{\epsilon_0}$ and $\ket{\epsilon_1}$ are the ground and excited states, respectively.
The steady-state density matrix can be calculated as
\begin{equation}
\varrho^{\rm ss}=\begin{pmatrix}
	\varrho_{00}^{\rm ss} & \varrho_{01}^{\rm ss}\\
	\varrho_{10}^{\rm ss} & \varrho_{11}^{\rm ss}
\end{pmatrix}=\begin{pmatrix}
	\dfrac{(n+1)[4\Delta^2+\gamma^2(2n+1)^2]+4(2n+1)\Omega^2}{(2n+1)(4\Delta^2 +\gamma^2(2n+1)^2+8\Omega^2)} & \dfrac{2\Omega[-2\Delta + i\gamma(2n+1)]}{(2n+1)(4\Delta^2 +\gamma^2(2n+1)^2+8\Omega^2)}\\
	\dfrac{2\Omega[-2\Delta - i\gamma(2n+1)]}{(2n+1)(4\Delta^2 +\gamma^2(2n+1)^2+8\Omega^2)} & \dfrac{n[4\Delta^2+\gamma^2(2n+1)^2]+4(2n+1)\Omega^2}{(2n+1)(4\Delta^2 +\gamma^2(2n+1)^2+8\Omega^2)}
\end{pmatrix},
\end{equation}
where $\varrho_{ij}^{\rm ss}=\mel{\epsilon_i}{\varrho^{\rm ss}}{\epsilon_j}$.
Using the energy eigenbasis, the matrix representations of the Lindblad super-operator is given by
\begin{align}
\hat{\mca{L}}&=\begin{pmatrix}
	-\gamma n & i\Omega & -i\Omega & \gamma(n+1) \\
	i\Omega & i\Delta-\gamma(2n+1)/2 & 0 & -i\Omega\\
	-i\Omega & 0 & -i\Delta-\gamma(2n+1)/2 & i\Omega\\
	\gamma n & -i\Omega & i\Omega & -\gamma(n+1)
\end{pmatrix}.
\end{align}
By performing some algebraic calculations, the explicit forms of the upper bounds $\mca{Q}_1^u$ and $\mca{Q}_2^u$ are obtained,
\begin{align}
\mca{Q}_1^u&=\frac{\mca{N}_1^u}{\mca{D}_1^u},\\
\mca{Q}_2^u&=\frac{\mca{N}_2^u}{\mca{D}_2^u},
\end{align}
where $\mca{N}_i^u$ and $\mca{D}_i^u$ can be written in terms of the parameters $\Omega$, $\Delta$, $\gamma$, $x=n(n+1)$, and $y=2n+1$ as
\begin{align}
\mca{N}_1^u&= 8\tau\left[32\Omega^4\left(\gamma^2y^4+2\Delta^2(8x+1)\right)+2\Omega^2\left(\gamma^2\Delta ^2(36x+1)y^2+\gamma^4y^6+4\Delta^4(20x+1)\right)+x\left(4\Delta^3+\Delta\gamma^2y^2\right)^2+128y^2\Omega^6\right]\notag \\
&\times\left[16x\Omega^2\left(4\Delta^2+\gamma^2y^2\right)+x\left(4 \Delta ^2+\gamma^2y^2\right)^2+16y^2\Omega^4\right],\\
\mca{D}_1^u&=\gamma y^3 \left[x\left(4\Delta^2+\gamma^2y^2\right)+2\Omega^2y^2\right]\left[4 \left(\Delta^2+2\Omega^2\right)+\gamma^2y^2\right]^3,\\
\mca{N}_2^u&=8\tau\big[16\Omega^4 \left(\gamma^2\Delta^2y^2(100x+1)+\gamma^4y^4(12x+1)+4 \Delta^4(52x+1)\right)+256\Omega^6\left(\gamma^2(6x+1)y^2+2\Delta^2(12x+1)\right)\notag\\
&+8x\Omega^2 \left(4\Delta^2+\gamma^2y^2\right)^2\left(6\Delta^2+\gamma^2y^2\right)+\Delta^2x\left(4 \Delta^2+\gamma^2y^2\right)^3+1024y^2\Omega^8\big],\\
\mca{D}_2^u&=\gamma y^3\left[4\left(\Delta^2+2\Omega^2\right)+\gamma^2y^2\right]^3.
\end{align}
Notice that both $\mca{Q}_1^u$ and $\mca{Q}_2^u$ converge to the same value when $\Omega\to 0$ or $n\to 0$.

\section{Sufficient conditions for the survival of the classical uncertainty relations}
Here we reveal the deterministic role of quantum coherence in constraining the precision of observables and derive sufficient conditions for the classical uncertainty relations to survive.
As stated in the main text, we consider the generic case of dissipative processes where the Hamiltonian has no energy degeneracy and the jump operators account for transitions between energy eigenstates with the same energy change.
Specifically, they satisfy $[L_k,H]=\omega_kL_k$, where $\omega_k$ denotes the energy change.
As a consequence, we have $[H,L_k^\dagger L_k]=0$.
Suppose that $H=\sum_n\epsilon_n\dyad{\epsilon_n}$, where $\epsilon_n$ is the eigenvalue and $\ket{\epsilon_n}$ is the corresponding eigenstate, then $\epsilon_m\neq\epsilon_n$ for all $m\neq n$.
Each jump operator thus has the form $L_k=\sum_{k^-}\gamma_{k,k^-}\dyad{\epsilon_{k^+}}{\epsilon_{k^-}}$, where $\gamma_{k,k^-}$ are complex numbers.

In the following, we show that $q_i(\Gamma_\tau)\le 0~(i=1,2,3)$ for arbitrary trajectory $\Gamma_\tau$.
Since $H_\theta$ and $L_{k,\theta}^\dagger L_{k,\theta}$ are diagonal in the eigenbasis of the Hamiltonian, we have
\begin{align}
U_{\theta}(t_{j+1},t_j)\ket{\epsilon_n}&=e^{u_n(\theta,t_{j+1},t_j)}\ket{\epsilon_n},
\end{align}
where $u_n(\theta,t_{j+1},t_j)$ is a complex function that is linear in $\theta$.
Because the initial pure state $\ket{n}$ can be expressed in terms of eigenstates as $\ket{n}=\sum_mc_{nm}\ket{\epsilon_m}$, the quantity $q_i(\Gamma_\tau)$ can be expanded in terms of the transition paths between eigenstates as
\begin{equation}\label{eq:path.expand}
\begin{aligned}[b]
q_i(\Gamma_\tau)&=-\left.\partial_\theta^2\ln|U_{\theta}(\tau,t_N)\prod_{j=1}^{N}L_{k_{j}}U_{\theta}(t_{j},t_{j-1})\ket{n}|^2\right|_{\theta=0}\\
&=-\left.\partial_\theta^2\ln|\sum_{m,\Gamma_{\tau}^p}\delta(m,\Gamma_{\tau}^p) e^{a(m,\Gamma_{\tau}^p)}\ket{\epsilon_{k_N^+}}|^2\right|_{\theta=0}.
\end{aligned}
\end{equation}
In Eq.~\eqref{eq:path.expand}, each path $\Gamma_{\tau}^p$ is described by a set of pairs of indices $\Gamma_{\tau}^p=\{(k_1^-,k_1^+),(k_2^-,k_2^+),\dots,(k_N^-,k_N^+)\}$.
Accordingly, the functional $\delta(m,\Gamma_{\tau}^p)$ and action $a(m,\Gamma_{\tau}^p)$ are defined as follows:
\begin{align}
\delta(m,\Gamma_{\tau}^p)&=\delta_{m,k_1^-}\delta_{k_1^+,k_2^-}\dots\delta_{k_{N-1}^+,k_N^-},\\
a(m,\Gamma_{\tau}^p)&=b(m,\Gamma_{\tau}^p)+u_{m}(\theta,t_{1},0)+\sum_{j=1}^{N}u_{k_j^+}(\theta,t_{j+1},t_j),
\end{align}
where $\delta_{m,n}$ is the Kronecker delta, $t_{N+1}\coloneqq\tau$, and $b(m,\Gamma_{\tau}^p)$ is a $\theta$-independent functional.
For each index $m$, there exists at most one path $\Gamma_{\tau}^p$ such that $\delta(m,\Gamma_{\tau}^p)=1$ and $k_N^+$ is different in each path.
By defining $a(m,\Gamma_{\tau}^p)+a(m,\Gamma_{\tau}^p)^*\eqqcolon \theta x(m,\Gamma_{\tau}^p)+y(m,\Gamma_{\tau}^p)\eqqcolon z(m,\Gamma_{\tau}^p)$, where $x$, $y$, and $z$ are real functionals, we have
\begin{equation}\label{eq:path.expand.2}
\begin{aligned}[b]
q_i(\Gamma_\tau)&=-\eval{\partial_\theta^2\ln\qty[\sum_{m,\Gamma_{\tau}^p}\delta(m,\Gamma_{\tau}^p) |e^{a(m,\Gamma_{\tau}^p)}|^2]}_{\theta=0}\\
&=-\eval{\partial_\theta^2\ln\qty[\sum_{m,\Gamma_{\tau}^p}\delta(m,\Gamma_{\tau}^p) e^{a(m,\Gamma_{\tau}^p)+a(m,\Gamma_{\tau}^p)^*}]}_{\theta=0}\\
&=-\eval{\partial_\theta^2\ln\qty[\sum_{m,\Gamma_{\tau}^p}\delta(m,\Gamma_{\tau}^p) e^{\theta x(m,\Gamma_{\tau}^p)+ y(m,\Gamma_{\tau}^p)}]}_{\theta=0}\\
&=\frac{\qty[\sum_{m,\Gamma_{\tau}^p}\delta(m,\Gamma_{\tau}^p)x(m,\Gamma_{\tau}^p)e^{z(m,\Gamma_{\tau}^p)}]^2-\sum_{m,\Gamma_{\tau}^p}\delta(m,\Gamma_{\tau}^p)e^{z(m,\Gamma_{\tau}^p)}\sum_{m,\Gamma_{\tau}^p}\delta(m,\Gamma_{\tau}^p)x(m,\Gamma_{\tau}^p)^2e^{z(m,\Gamma_{\tau}^p)}}{\qty[\sum_{m,\Gamma_{\tau}^p}\delta(m,\Gamma_{\tau}^p) e^{z(m,\Gamma_{\tau}^p)}]^2}.
\end{aligned}
\end{equation}
According to the Cauchy--Schwarz inequality, we have
\begin{equation}
\qty[\sum_{m,\Gamma_{\tau}^p}\delta(m,\Gamma_{\tau}^p)x(m,\Gamma_{\tau}^p)e^{z(m,\Gamma_{\tau}^p)}]^2-\sum_{m,\Gamma_{\tau}^p}\delta(m,\Gamma_{\tau}^p)e^{z(m,\Gamma_{\tau}^p)}\sum_{m,\Gamma_{\tau}^p}\delta(m,\Gamma_{\tau}^p)x(m,\Gamma_{\tau}^p)^2e^{z(m,\Gamma_{\tau}^p)}\le 0.
\end{equation}
Therefore, $q_i(\Gamma_\tau)\le 0$ and thus, $\mca{Q}_i\le 0$ for all $i=1,2,3$. 
Consequently, the classical KURs survive in the quantum regime,
\begin{align}
\frac{\Var{O}}{\expval{O}^2}&\ge\frac{(1+\delta O)^2}{\mca{A}_\tau},~\text{for}~O\in\{\Phi,\Lambda\},\\
\frac{\Var{\tau}}{\expval{\tau}^2}&\ge\frac{1}{\expval{N}_\tau}.
\end{align}

In the case of nonresonant processes, that is, $L_k=\sqrt{\gamma_{mn}}\dyad{\epsilon_m}{\epsilon_n}$ for all $k$, we show that $\tilde\delta J=\delta J$.
To this end, we first have
\begin{align}
\dot\varrho_t&=\mca{L}(\varrho_t),\\
\dot\phi_t&=\mca{L}(\varrho_t+\phi_t)+\sum_k[\ell_k(t)-1]\mca{D}[L_k]\varrho_t,
\end{align}
with $\phi_0=\mbb{0}$.
Defining $p_n(t)\coloneqq\mel{\epsilon_n}{\varrho_t}{\epsilon_n}$ and $q_n(t)\coloneqq\mel{\epsilon_n}{\phi_t}{\epsilon_n}$, we obtain the time evolution of these population distributions as follows:
\begin{align}
\dot p_n(t)&=\sum_{m(\neq n)}[\gamma_{nm}p_m(t)-\gamma_{mn}p_n(t)],\label{eq:master.eq1} \\
\dot q_n(t)&=\sum_{m(\neq n)}[\gamma_{nm}(p_m(t)+q_m(t))-\gamma_{mn}(p_n(t)+q_n(t))] \label{eq:master.eq2}.
\end{align}
By defining $\ket{p_t}\coloneqq[p_n(t)]^\top$, $\ket{q_t}\coloneqq[q_n(t)]^\top$, and $\msf{R}\coloneqq [r_{mn}]$ with $r_{mn}=\gamma_{mn}$ for $m\neq n$ and $r_{nn}=-\sum_{m(\neq n)}r_{mn}$, Eqs.~\eqref{eq:master.eq1} and \eqref{eq:master.eq2} can be rewritten as follows:
\begin{align}
\ket{\dot p_t}&=\msf{R}\ket{p_t},\\
\ket{\dot q_t}&=\msf{R}(\ket{p_t}+\ket{q_t}).
\end{align}
The solution of these equations can be explicitly calculated as
\begin{align}
\ket{p_t}&=e^{\msf{R}t}\ket{p_0},\\
\ket{q_t}&=t\msf{R}e^{\msf{R}t}\ket{p_0}+e^{\msf{R}t}\ket{q_0}=t\ket{\dot p_t}.
\end{align}
Here, we have used $\msf{R}e^{\msf{R}t}\ket{p_0}=\ket{\dot p_t}$ and $\ket{q_0}=0$ in the last equality.
Using these relations, we can calculate as follows:
\begin{equation}
\begin{aligned}[b]
\expval{J}+\expval{J}_*&=\int_0^\tau\sum_kw_k\tr{L_k(\varrho_t+\phi_t)L_k^\dagger}\dd{t}\\
&=\int_0^\tau\sum_kw_k\gamma_{mn}(p_n(t)+q_n(t))\dd{t}\\
&=\int_0^\tau\bra{w}(\ket{p_t}+\ket{q_t})\dd{t}\\
&=\int_0^\tau\bra{w}(\ket{p_t}+t\ket{\dot p_t})\dd{t}\\
&=\int_0^\tau\frac{d}{dt}[t\braket{w}{p_t}]\dd{t}\\
&=\tau\braket{w}{p_\tau}\\
&=\tau\sum_kw_k\tr{L_k\varrho_\tau L_k^\dagger}\\
&=\tau\partial_\tau\expval{J}.
\end{aligned}
\end{equation}
Consequently, $\tilde\delta J=\expval{J}_*/\expval{J}=\tau\partial_\tau\ln|\expval{J}|-1=\tau\partial_\tau\ln|\expval{J}/\tau|=\delta J$.
Thus, the classical TUR survives in this case as $\mca{Q}_1\le 0$,
\begin{equation}
\frac{\Var{J}}{\expval{J}^2}\ge\frac{2(1+\delta J)^2}{\Sigma_\tau}.
\end{equation}

\section{Thermodynamics of the three-level maser}

Here we describe the thermodynamics of the three-level maser employed in the illustrative example.
The dynamics of the density matrix is governed by the local master equation
\begin{equation}\label{eq:local.mas.eq}
\dot\varrho_t=-i[H_t,\varrho_t]+\sum_{k=1}^2\qty(\mca{D}[L_{k}]\varrho_t+ \mca{D}[L_{k'}]\varrho_t),
\end{equation}
where the Hamiltonian and jump operators are given by
\begin{align}
H_t&=H_0+V_t,\\
L_1&= \sqrt{\gamma_hn_h}\sigma_{31},\\
L_{1'}&=\sqrt{\gamma_h(n_h+1)}\sigma_{13},\\
L_2&=\sqrt{\gamma_cn_c}\sigma_{32},\\
L_{2'}&=\sqrt{\gamma_c(n_c+1)}\sigma_{23}.
\end{align}
Here, $H_0=\qty( \omega_1\sigma_{11} + \omega_2\sigma_{22} + \omega_3\sigma_{33} )$ is the bare Hamiltonian and $V_t=\Omega\qty( e^{i\omega_0t}\sigma_{12}+ e^{-i\omega_0t}\sigma_{21}) $ is the external classical field.
To remove the time dependence of the full Hamiltonian, it is convenient to rewrite operators in the rotating frame $X\to\tilde{X}= U_t^\dagger XU_t$, where $U_t=e^{-i\bar{H}t}$ and $\bar{H}=\omega_1\sigma_{11}+(\omega_1+\omega_0)\sigma_{22}+\omega_3\sigma_{33}$.
In this rotating frame, the master equation reads
\begin{equation}\label{eq:rot.frame.Ham}
\dot{\tilde{\varrho}}_t = -i[ H,\tilde{\varrho}_t ] +\sum_{k=1}^2\qty(\mca{D}[L_{k}]\tilde{\varrho}_t+ \mca{D}[L_{k'}]\tilde{\varrho}_t),
\end{equation}
where $H=-\Delta \sigma_{22}+\Omega( \sigma_{12}+\sigma_{21} )$ and $\Delta=\omega_0+\omega_1-\omega_2$.
Equation \eqref{eq:rot.frame.Ham} is exactly the master equation considered in the main text.
It was shown that the master equation \eqref{eq:local.mas.eq} is valid when the driving field is weak \cite{Geva.1996.JCP}.
In the case of strong driving fields, the local master equation should be modified to be thermodynamically consistent \cite{Geva.1996.JCP}.
In the present paper, we exclusively consider the case of weak driving fields, and the validity of Eq.~\eqref{eq:local.mas.eq} is thus guaranteed.

Now we consider the thermodynamics of the three-level maser described by Eq.~\eqref{eq:local.mas.eq}.
Following the approach proposed in Ref.~\cite{Boukobza.2007.PRL}, the first law of thermodynamics can be formulated as
\begin{equation}\label{eq:first.law}
\begin{aligned}[b]
\dot{E}_t\coloneqq\frac{d}{dt}\tr{\varrho_tH_0}=\tr{\dot\varrho_tH_0}&=\tr{\sum_{k=1}^2\qty(\mca{D}[L_{k}]\varrho_t+ \mca{D}[L_{k'}]\varrho_t)H_0}-i\tr{[H_0,V_t]\varrho_t}\\
&\eqqcolon\dot{Q}_t+\dot{W}_t,
\end{aligned}
\end{equation}
where $\dot{E}_t$ is the energy change and $\dot{Q}_t$ and $\dot{W}_t$ denote the heat and work flux, respectively.
The heat flux can be decomposed into two contributions from the hot and cold heat baths,
\begin{equation}
\begin{aligned}[b]
\dot{Q}_t&=\tr{\sum_{k=1}^2\qty(\mca{D}[L_{k}]\varrho_t+ \mca{D}[L_{k'}]\varrho_t)H_0}=\tr{\qty(\mca{D}[L_{1}]\varrho_t+ \mca{D}[L_{1'}]\varrho_t)H_0}+\tr{\qty(\mca{D}[L_{2}]\varrho_t+ \mca{D}[L_{2'}]\varrho_t)H_0}\\
&\eqqcolon \dot{Q}_t^{(h)} + \dot{Q}_t^{(c)}.
\end{aligned}
\end{equation}
Consequently, the entropy production rate reads
\begin{equation}
\dot{\Sigma}_t=\dot{S}_t-\frac{\dot{Q}_t^{(h)}}{T_h}-\frac{\dot{Q}_t^{(c)}}{T_c}\ge 0.
\end{equation}
Here $S_t\coloneqq -\tr{\varrho_t\ln\varrho_t}$ denotes the von Neumann entropy.
The nonnegativity of the entropy production rate corresponds to the second law of thermodynamics.
Note that the entropy production does not change in the rotating frame.
Since the environmental entropy changes due to the jumps are $\Delta s_{1'}=-\Delta s_{1}=\beta_h(\omega_3-\omega_1)$ and $\Delta s_{2'}=-\Delta s_{2}=\beta_c(\omega_3-\omega_2)$, and the von Neumann entropy is invariant under unitary transforms, we have
\begin{equation}
\begin{aligned}
\Sigma_\tau&=-\tr{\varrho_\tau\ln\varrho_\tau}+\tr{\varrho_0\ln\varrho_0}\\
&+\int_0^\tau\qty[ \beta_h(\omega_3-\omega_1)(\tr{L_{1'}\varrho_tL_{1'}^\dagger}-\tr{L_1\varrho_tL_1^\dagger}) + \beta_c(\omega_3-\omega_2)(\tr{L_{2'}\varrho_tL_{2'}^\dagger}-\tr{L_2\varrho_tL_2^\dagger}) ]\dd{t}\\
&=-\tr{\tilde\varrho_\tau\ln\tilde\varrho_\tau}+\tr{\tilde\varrho_0\ln\tilde\varrho_0}+\int_0^\tau\qty[ \Delta s_{1'}(\tr{L_{1'}\tilde\varrho_tL_{1'}^\dagger}-\tr{L_1\tilde\varrho_tL_1^\dagger}) + \Delta s_{2'}(\tr{L_{2'}\tilde\varrho_tL_{2'}^\dagger}-\tr{L_2\tilde\varrho_tL_2^\dagger}) ]\dd{t}\\
&=-\tr{\tilde\varrho_\tau\ln\tilde\varrho_\tau}+\tr{\tilde\varrho_0\ln\tilde\varrho_0}+\int_0^\tau\sum_k\Delta s_{k}(\tr{L_{k}\tilde\varrho_tL_{k}^\dagger}\dd{t}\\
&=\Delta S_{\rm sys} + \Delta S_{\rm env},
\end{aligned}
\end{equation}
which is exactly the formula of the irreversible entropy production defined in the main text.
It is also worth noting that if we use the full Hamiltonian $H_t$ instead of the bare Hamiltonian $H_0$ in Eq.~\eqref{eq:first.law}, the corresponding entropy production can be negative.
An approach to resolve this issue is to modify the local master equation by adding a correction term \cite{Geva.1996.JCP}.